\documentclass[10pt,journal,compsoc]{IEEEtran}
\usepackage{lineno,hyperref}
\usepackage{times}
\usepackage{latexsym}
\usepackage{graphicx,subfigure}
\usepackage{epstopdf}
\usepackage{amsmath, amsfonts}
\usepackage{breqn, algorithm, algpseudocode}
\usepackage{multirow}
\usepackage{upgreek}
\usepackage{bm}
\usepackage{cite}
\usepackage{indentfirst}

\begin{document}

\title{Short Text Topic Modeling Techniques, Applications, and Performance: A Survey}

\author{Jipeng~Qiang, Zhenyu~Qian, Yun~Li, Yunhao~Yuan, and~Xindong~Wu, ~\IEEEmembership{~Fellow,~IEEE,}
\IEEEcompsocitemizethanks{\IEEEcompsocthanksitem J. Qiang, Z. Qian, Y. Li and Y. Yuan are with the Department
of Computer Science, Yangzhou, Jiangsu, P. R. China, 225127.\protect\\
E-mail: \{jpqiang,liyun, yhyuan\}@yzu.edu.cn
\IEEEcompsocthanksitem X. Wu is with Department of Computer Science, University of Louisiana at Lafayette, Louisiana, USA. \protect\\
E-mail: xwu@louisiana.edu.edu
}
}

\markboth{Journal of \LaTeX\ Class Files,~Vol.~14, No.~8, April~2019}%
{Shell \MakeLowercase{\textit{et al.}}: Bare Advanced Demo of IEEEtran.cls for IEEE Computer Society Journals}

\IEEEtitleabstractindextext{%
\begin{abstract}
Analyzing short texts infers discriminative and coherent latent topics that is a critical and fundamental task since many real-world applications require semantic understanding of short texts. Traditional long text topic modeling algorithms (e.g., PLSA and LDA) based on word co-occurrences cannot solve this problem very well since only very limited word co-occurrence information is available in short texts. Therefore, short text topic modeling has already attracted much attention from the machine learning research community in recent years, which aims at overcoming the problem of sparseness in short texts. In this survey, we conduct a comprehensive review of various short text topic modeling techniques proposed in the literature. We present three categories of methods based on Dirichlet multinomial mixture, global word co-occurrences, and self-aggregation, with example of representative approaches in each category and analysis of their performance on various tasks. We develop the first comprehensive open-source library, called STTM, for use in Java that integrates all surveyed algorithms within a unified interface, benchmark datasets, to facilitate the expansion of new methods in this research field. Finally, we evaluate these state-of-the-art methods on many real-world datasets and compare their performance against one another and versus long text topic modeling algorithm. 
\end{abstract}

}

\maketitle

\IEEEdisplaynontitleabstractindextext

\IEEEpeerreviewmaketitle

\section{Introduction}

Short texts have become an important information source including news headlines, status updates, web page snippets, tweets, question/answer pairs, etc.  Short text analysis has been attracting increasing attention in recent years due to the ubiquity of short text in the real-world \cite{Lin2014The,qiang2016topic,shi2018short}. Effective and efficient models infer the latent topics from short texts, which can help discover the latent semantic structures that occur in a collection of documents. Short text topic modeling algorithms are always applied into many tasks such as topic detection \cite{wang2007mining}, classification \cite{sriram2010short}, comment summarization \cite{ma2012topic}, user interest profiling \cite{weng2010twitterrank}.

Traditional topic modeling algorithms such as probabilistic latent semantic analysis (PLSA) \cite{hofmann1999probabilistic} and latent Dirichlet allocation (LDA) \cite{blei2003latent} are widely adopted for discovering latent semantic structure from text corpus without requiring any prior annotations or labeling of the documents. In these algorithms, each document may be viewed as a mixture of various topics and each topic is characterized by a distribution over all the words. Statistical techniques (e.g., Variational methods and Gibbs sampling) are then employed to infer the latent topic distribution of each document and the word distribution of each topic using higher-order word co-occurrence patterns \cite{blei2012probabilistic}. These algorithms and their variants have had a major impact on numerous applied fields in modeling text collections news articles, research papers, and blogs \cite{Hoffman2010Online,xie2013integrating,xie2015incorporating}. However, traditional topic models experience large performance degradation over short texts due to the lack of word co-occurrence information in each short text \cite{cheng2014btm, Lin2014The}. Therefore, short text topic modeling has already attracted much attention from the machine learning research community in recent years, which aims at overcoming the problem of sparseness in short texts.

Earlier works \cite{jin2011transferring,phan2008learning} still used traditional topic models for short texts, but exploited external knowledge or metadata to bring in additional useful word co-occurrences across short texts, and therefore may boost the performance of topic models. For example, Phan et al. \cite{phan2008learning} first learned latent topics from Wikipedia, and then inferred topics from short texts. Weng et al. \cite{weng2010twitterrank} and Mehrotra et al. \cite{mehrotra2013improving} aggregated tweets for pseudo-document using hashtags and the same user respectively. The problem lies in that auxiliary information or metadata is not always available or just too costly for deployment. These studies suggest that topic models specifically designed for general short texts are imperative. This survey will provide a taxonomy that captures the existing short text topic modeling algorithms and their application domains.

\begin{table*}
  \centering
  \caption{An event about artificial intelligence was reported by different news media on March 1, 2018.} \label{example}
  \begin{tabular}{|c|l|l|} \hline
  \textbf{Number} & \textbf{Media} & \textbf{Headline} \\ \hline
  $1$ & Lawfare & President Trump's Executive Order on Artificial Intelligence\\\hline
  $2$ &Nextgov & White House’s Race to Maintain AI Dominance Misses Opportunity\\\hline
  $3$& Forbes & Artificial Intelligence Regulation may be Impossible \\\hline
  $4$&  CognitiveWorld &Pop Culture, AI and Ethics \\
  \hline
  
  \end{tabular}
\end{table*}

News aggregation websites often rely on news headlines to cluster different source news about the same event. In Table \ref{example}, we show an event about artificial intelligence reported on March 1, 2018. As presented, all these short texts were reported about the same event. From these short texts, we can found these following characteristics. (1) Obviously, each short text lacks enough word co-occurrence information. (2) Due to a few words in each text, most texts are probably generated by only one topic (e.g, text 1, text2, text 3). (3) Statistical information of words among texts cannot fully capture words that are semantically related but rarely co-occur. For example, President Trump of text 1 and White House of text 2 are highly semantically related, and AI is short for Artifical Intelligence. (4) The single-topic assumption may be too strong for some short texts. For example, text 3 is probably associated with a small number of topics (e.g., one to three topics). Considering these characteristics, existing short text topic modeling algorithms were proposed by trying to solve one or two of these characteristics. Here, we divide the short text topic modeling algorithms basically into the following three major categories.

\textbf {(1) Dirichlet multinomial mixture (DMM) based methods}: A simple and effective model, Dirichlet Multinomial Mixture model, has been adopted to infer latent topics in short texts \cite{yin2014dirichlet,zhao2011comparing}. DMM follows the simple assumption that each text is sampled from only one latent topic. Considering the characteristics (1) and (2) in short texts, this assumption is reasonable and suitable for short texts compared to the complex assumption adopted by LDA that each text is modeled over a set of topics \cite{quan2015short,yan2015probabilistic}. Nigam et al. \cite{nigam2000text} proposed an EM-based algorithm for Dirichlet Multinomial Mixture (DMM) model. Except for the basic expectation maximization (EM), a number of inference methods have been used to estimate the parameters including variation inference and Gibbs sampling. For example, Yu et al. \cite{yu2010document} proposed the DMAFP model based on variational inference algorithm \cite{huang2013dirichlet}. Yin et al. \cite{yin2014dirichlet} proposed a collapsed Gibbs sampling algorithm for DMM. Other variations based on DMM \cite{qiang2018short,yin2016text,yin2018text} were proposed for improving the performance. The above models based on DMM ignore the characteristic (3). Therefore, many models by incorporating word embeddings into DMM were proposed \cite{nguyen2015improving,li2016topic}, because word embeddings learned from millions of external documents contain semantic information of words \cite{Mikolov2013Dis}. Not only word co-occurrence words belong to one topic, but words with high similarity have high probability belonging to one topic, which can effectively solve the data sparsity issue. To highlight the characteristic (4), a Poisson-based DMM model (PDMM) was proposed that allows each short text is sampled by a limited number of topics \cite{li2017enhancing}. Accordingly, Li et al. \cite{li2017enhancing} proposed a new model by directly extending the PDMM model using word embeddings.

\textbf{(2) Global word co-occurrences based methods}: Considering the characteristic (1), some models try to use the rich global word co-occurrence patterns for inferring latent topics \cite{cheng2014btm,zuo2016word}. Due to the adequacy of global word co-occurrences, the sparsity of short texts is mitigated for these models. According to the utilizing strategies of global word co-occurrences, this type of models can be divided into two types. 1) The first type directly uses the global word co-occurrences to infer latent topics. Biterm topic modeling (BTM) \cite{cheng2014btm} posits that the two words in a biterm share the same topic drawn from a mixture of topics over the whole corpus. Some models extend the Biterm Topic Modeling (BTM) by incorporating the burstiness of biterms as prior knowledge \cite{yan2015probabilistic} or distinguishing background words from topical words \cite{chen2015user}. 2) The second type first constructs word co-occurrence network using global word co-occurrences and then infers latent topics from this network, where each word correspond to one node and the weight of each edge stands for the empirical co-occurrence probability of the connected two words \cite{zuo2016word, wang2017learning}. 

\textbf{(3) Self-aggregation based methods}: Self-aggregation based methods are proposed to perform topic modeling and text self-aggregation during topic inference simultaneously. Short texts are merged into long pseudo-documents before topic inference that can help improve word co-occurrence information. Different from the aforementioned aggregation strategies \cite{weng2010twitterrank,mehrotra2013improving}, this type of methods SATM \cite{quan2015short} and PTM\cite{zuo2016topic} posit that each short text is sampled from a long pseudo-document unobserved in current text collection, and infer latent topics from long pseudo-documents, without depending on auxiliary information or metadata. Considering the characteristic (3), Qiang et al. \cite{qiang2017topic} and Bicalho et al. \cite{bicalho2017general} merged short texts into long pseudo-documents using word embeddings.

\subsection{Our contributions}

This survey has the following three-pronged contribution:

(1) We propose a taxonomy of algorithms for short text topic modeling and explain their differences. We define three different tasks, i.e., application domains of short text topic modeling techniques. We illustrate the evolution of the topic, the challenges it faces, and future possible research directions.

(2) To facilitate the expansion of new methods in this field, we develop the first comprehensive open-source JAVA library, called STTM, which not only includes all short text topic modeling algorithms discussed in this survey with a uniform easy-to-use programming interface but also includes a great number of designed modules for the evaluation and application of short text topic modeling algorithms. STTM is open-sourced at https://github.com/qiang2100/STTM.

(3) We finally provide a detailed analysis of short text topic modeling techniques and discuss their performance on various applications. For each model, we analyze their results through comprehensive comparative evaluation on some common datasets.

\subsection{Organization of the survey}

The rest of this survey is organized as follows. In Section 2, we introduce the task of short text topic modeling. Section 3 proposes a taxonomy of short text topic modeling algorithms and provides a description of representative approaches in each category. The list of applications for which researchers have used the short text topic modeling algorithms is provided in Section 4. Section 5 presents our Java library for short text topic modeling algorithms. In the next two sections, we describe the experimental setup (Section 6) and evaluate the discussed models (Section 7).  Finally, we draw our conclusions and discuss potential future research directions in Section 8.

\section{Definitions}

In this section, we formally define the problem of short text topic modeling. 

Given a short text corpus $\bm{D}$ of $N$ documents, with a vocabulary $W$ of size $V$, and $K$ pre-defined latent topics. One document $d$ is represented as $(w_{d,1},w_{d,2},...,w_{d,n_d})$ in $D$ including $n_d$ words.

A topic $\phi$ in a given collection $\bm{D}$ is defined as a multinomial distribution over the vocabulary $W$, i.e., $\{p(w|\phi)\}_{w\in W}$. The topic representation of a document $d$, $\theta_d$, is defined as a multinomial distribution over $K$ topics, i.e., $\{p(\phi_k|\theta_d)\}_{k=1,...,K}$. The general task of topic modeling aims to find $K$ salient topics $\phi_{k=1,...,K}$ from $\bm{D}$ and to find the topic representation of each document $\theta_{d=1,...,N}$. 

Most classical probabilistic topic models adopt the Dirichlet prior for both the topics and the topic representation of documents, which are first used in LDA \cite{blei2003latent}, which is $\phi_k \sim Dirichlet(\beta)$ and $\theta_d \sim Dirichlet(\alpha)$. In practice, the Dirichlet prior smooths the topic mixture in individual documents and the word distribution of each topic, which alleviates the overfitting problem of probabilistic latent semantic analysis (PLSA) \cite{hofmann1999probabilistic}, especially when the number of topics and the size of vocabulary increase. Therefore, all of existing short text topic modeling algorithms adopt Dirichlet distribution as prior distribution. 

Given a short text corpus $\bm{D}$ with a vocabulary of size $V$, and the predefined number of topics $K$, the major tasks of short text topic modeling can be defined as to:

(1). Learn the word representation of topics $\phi$;

(2). Learn the sparse topic representation of documents $\theta$.

All the notations used in this paper are summarized in Table \ref{symbols}.

\begin{table}
  \centering
  \caption{The notations of symbols used in the paper} \label{symbols}
  \begin{tabular}{|c|l|} \hline
  $\bm{D},N$& Documents and number of documents in the corpus\\\hline
  $W,V$ & The vocabulary and number of words in the vocabulary\\\hline
  $K$& Number of pre-defined latent topics \\\hline
  $\overline{l}$& Average length of each document in $D$\\\hline
  $n_{k}$&  Number of words associated with topic $k$\\\hline
  $m_{k}$&  Number of documents associated with topic $k$\\\hline
  $n_{k}^w$&  Number of word $w$ associated with topic $k$ in \overrightarrow{d}\\\hline
  $n_d$&  Number of words in document $d$\\\hline
  $n_{d}^w$&  Number of word $w$ in document $d$\\\hline
  $n_{d}^k$&  Number of word associated with topic $k$ in document $d$\\\hline
  $n_{k,d}^{w}$&  Number of word $w$ associated with topic $k$ in document $d$\\\hline
  $P$&  Long pseudo-document set generated by models\\\hline
  $\phi$& Topic distribution\\\hline
  $\theta$& Document-topic distribution\\\hline
  $z$&  Topic indicator\\\hline
  $U$&  Number of dimensions in word embeddings\\\hline
  $\zeta$&  Time cost of considering GPU model\\\hline
  $\varsigma$&  Maximum number of topics allowable in a short text\\\hline
  $c$&  Size of sliding window \\\hline
  \end{tabular}
\end{table}

\section{Algorithmic Approaches: A Taxonomy}

In the past decade, there has been much work to discover latent topics from short texts using traditional topic modeling algorithms by incorporating external knowledge or metadata. More recently, researchers focused on proposing new short text topic modeling algorithms. In the following, we present historical context about the research progress in this domain, then propose a taxonomy of short text topic modeling techniques including: (1) Dirichlet Multinomial Mixture (DMM) based methods, (2) Global word co-occurrence based methods, and (3) Self-aggregation based methods.

\subsection{Short Text Topic Modeling Research Context and Evolution}

Traditional topic modeling algorithms such as probabilistic latent semantic analysis (PLSA) \cite{hofmann1999probabilistic} and latent Dirichlet allocation (LDA) \cite{blei2003latent} are widely adopted for discovering latent semantic structure from text corpus by capturing word co-occurrence pattern at the document level. Hence, more word co-occurrences would bring in more reliable and better topic inference. Due to the lack of word co-occurrence information in each short text, traditional topic models have a large performance degradation over short texts. Earlier works focus on exploiting external knowledge to help enhance the topic inference of short texts. For example, Phan et al. \cite{phan2008learning} adopted the learned latent topics from Wikipedia to help infer the topic structure of short texts. Similarly, Jin et al. \cite{jin2011transferring} searched auxiliary long texts for short texts to infer latent topics of short texts for clustering. A large regular text corpus of high quality is required by these models, which bring in big limitation for these models.

Since 2010, research on topic discovery from short texts has been shifted to merging short texts into long pseudo-documents using different aggregation strategies before adopting traditional topic modeling to infer the latent topics. For example, Weng et al. \cite{weng2010twitterrank} merge all tweets of one user into a pseudo-document before using LDA. Other information includes hashtags, timestamps, and named entities have been tread as metadata to merging short texts \cite{hong2010empirical,mehrotra2013improving,zhao2011comparing}. However, helpful metadata may not be accessible in any domains, e.g., news headlines and search snippets. These studies suggest that topic models specifically designed for general short texts are crucial. This survey will provide a taxonomy that captures the existing strategies and these application domains.

\subsection{A Taxonomy of Short Text Topic Modeling Methods}

We propose a taxonomy of short text topic modeling approaches. We categorize topic modeling approaches into three broad categories: (1) Dirichlet Multinomial Mixture (DMM) based, (2) Global Word co-occurrences based, and (3) Self-aggregation based. Below we describe the characteristics of each of these categories and present a summary of some representative methods for each category (cf. Table \ref{list}).

\begin{table*}
\centering
\caption{List of short text topic modeling approaches }\label{list}
\begin{tabular}{|c|c|c|c|c|c|} \hline
   \textbf{Category} & \textbf{Year} & \textbf{Published} & \textbf{Authors} & \textbf{Method} & \textbf{Time Complexity of One Iteration}  \\\hline
   \multirow{3}{*}{DMM} & 2014 & KDD \cite{yin2014dirichlet} & J. Yin \& et al. & GSDMM & $O(KN\overline{l})$\\
                       & 2015 & TACL \cite{nguyen2015improving} & D. Nguyen \& et al. & LF-DMM & $O(O(2KN\overline{l}+KVU))$\\
                       & 2016 & SIGIR \cite{li2016topic}& C. Li \& et al. & GPU-DMM & $O(KN\overline{l}+N\overline{l}\zeta+KV)$\\
                       & 2017 & TOIS \cite{li2017enhancing} & C. Li \& et al. & GPU-PDMM & $O(N\overline{l}\sum_{i=1}^{\varsigma-1}C_K^i+N\overline{l}\zeta+KV)$\\ \hline
   \multirow{2}{*}{Global word} & 2013 & WWW \cite{cheng2014btm}  & X. Chen \& et al. & BTM & $O(KN\overline{l}c)$ \\
                                     co-occurrences & 2016 & KAIS \cite{zuo2016word}  & Y. Zuo \& et al. & WNTM & $O(KN\overline{l}c(c-1))$\\ \hline
    \multirow{2}{*}{ Self-aggregation} & 2015 & IJCAI & X. Quan \& et al. & SATM & $O(N\overline{l}PK)$\\
                                       & 2016 & KDD & Y. Zuo \& et al. & PTM & $O(N\overline{l}(P+K))$ \\ \hline

  \hline\end{tabular}

\end{table*}

\subsection{Dirichlet Multinomial Mixture based Methods}

Dirichlet Multinomial Mixture model (DMM) was first proposed by Nigam et al. \cite{nigam2000text} based on the assumption that each document is sampled by only one topic. The assumption is more fit for short texts than the assumption that each text is generated by multiple topics. Therefore, many models for short texts were proposed based on this simple assumption \cite{qiang2018short,zhao2011comparing,yu2010document}. Yin et al. \cite{yin2014dirichlet} proposed a DMM model based on collapse Gibbs sampling. Zhao et al. \cite{zhao2011comparing} proposed a Twitter-LDA model by assuming that one tweet is generated from one topic. Recently, more work incorporates word embeddings into DMM \cite{li2017enhancing,li2016topic}.

\subsubsection{GSDMM}

\begin{figure}[H]
  \centering
  \includegraphics[width=80mm]{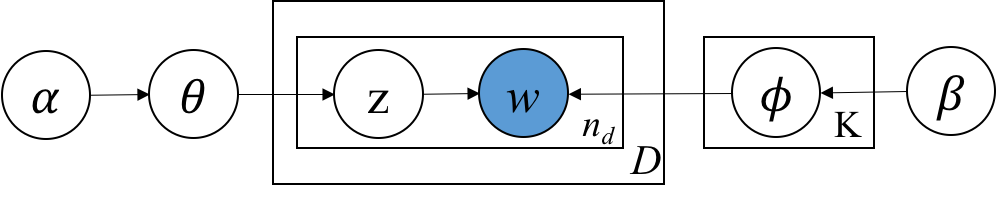}
  \caption{Graphical Model of GSDMM.}\label{gsdmm}
\end{figure}

DMM respectively chooses Dirichlet distribution for topic-word distribution $\phi$ and document-topic distribution $\theta$ as prior distribution with parameter $\alpha$ and $\beta$. DMM samples a topic $z_d$ for the document $d$ by Multinomial distribution $\theta$, and then generates all words in the document $d$ from topic $z_d$ by Multinomial distribution $\phi_{z_d}$. The graphical model of DMM is shown in Figure \ref{gsdmm}. The generative process for DMM is described as follows:

(1). Sample a topic proportion $\theta\sim Dirichlet(\alpha)$.

(2). For each topic $k\in\left\{1,...,K\right\}$:

    \qquad Draw a topic-word distribution $\theta_k\sim Dirichlet(\beta)$.

(3). For each document $d \in \bm{D}$:

    \qquad(b)Sample a topic $z_d \sim Multinomial(\theta)$.

    \qquad(c)For each word $w\in\left\{w_{d,1},...,w_{d,n_d}\right\}$:

           \qquad\qquad Sample a word $w\sim Multinomial(\phi_{z_d})$.

Gibbs sampling algorithm for Dirichlet Multinomial Mixture model is denoted as GSDMM, which is based on the assumption that each text is sampled by a single topic \cite{yin2014dirichlet}. Here, for better representation of latent topics, we represent a topic with the topic feature (CF) vector, which essentially is a big document combined with its documents. 

The TF vector of a topic $k$ is defined as a tuple $\{n_k^w (w\in W), m_k, n_k\}$, where $n_k^w$ is the number of word $w$ in topic $k$, $m_k$ is the number of documents in topic $k$, and $n_k$ is the number of words in topic $k$.

The topic feature (TF) presents important addible and deletable properties, as described next.

(1) \textbf{Addible Property}. A document $d$ can be efficiently added to topic $k$ by updating its TF vector as follows.

\[
\begin{split}
& n^w_k = n^w_k + n^w_d \ \ \ for \ each \ word \ w \ in \ d \\
& m_k = m_k + 1 \ ; \ n_k = n_k + n_d\\
 \end{split}
\]

(2) \textbf{Deletable Property}. A document $d$ can be efficiently deleted from topic $k$ by updating its TF vector as follows.

\[
  \begin{split}
& n^w_k = n^w_k - n^w_d \ \ \  for \  each \ word \ w \ in \ d \\
& m_k = m_k - 1 \ ; \ n_k = n_k - n_d\\
 \end{split}
\]

The hidden multinomial variable ($z_d$) for document $d$ is sampled based on collapsed Gibbs sampling, conditioned on a complete assignment of all other hidden variables. GSDMM uses the following conditional probability distribution to infer its topic,

\begin{equation} \label{equation1}
\begin{split}
 p(z_d=k| & \bm{Z}_{\neg{d}},\bm{D}) \propto \\
 &\frac{m_{k,\neg{d}}+\alpha}{N-1+K\alpha} \frac{\prod_{w\in d}\prod_{j=1}^{n_d^w}(n_{k,\neg{d}}^w+\beta+j-1)}{\prod_{i=1}^{n_{d}}(n_{k,\neg{d}}+V\beta+i-1)}\\
\end{split}
\end{equation}
where $\bm{Z}$ represents all topics of all documents, the subscript $\neg{d}$ means document $d$ is removed from its current topic feature (TF) vector, which is useful for the update learning process of GSDMM. 

For each document, we first delete it from its current TF vector with the deletable property. Then, we reassign the document to a topic according to the probability of the document belonging to each of the $K$ topics using Equation \ref{equation1}. After obtaining the topic of the document, we add it from its new TF vector with the addible property. Finally, the posterior distribution of each word belonging to each topic is calculated as the follows,

\begin{equation} \label{equation2}
\phi_k^w=\frac{n^w_k+\beta}{n_k+V\beta}
\end{equation}

\subsubsection{LF-DMM}

\begin{figure}[H]
  \centering
  \includegraphics[width=80mm]{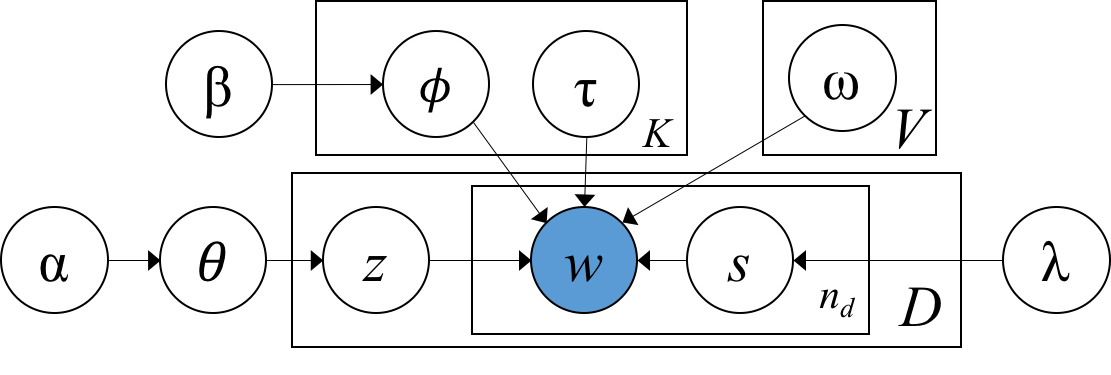}
  \caption{Graphical Model of LF-DMM.}\label{lfdmm}
\end{figure}

The graphical model of LF-DMM is shown in Figure \ref{lfdmm}. Based on the assumption that each text is sampled by a single topic, LF-DMM generates the words by Dirichlet multinomial model or latent feature model.  Given two latent-feature vectors $\uptau$ associated with topic $k$ and $\upomega$ associated with word $w$, latent feature model generates a word $w$ using $softmax$ function by the formula,

\begin{equation} \label{softmax}
\sigma(w\mid \uptau_k\bm{\upomega}^T)=\frac{e^{(\uptau_k\cdot\upomega_w)}}{\Sigma_{w'\in W}e^{(\uptau_k\cdot\upomega_{w'})}}
\end{equation}
where $\bm{\upomega}$ is pre-trained word vectors of all words $W$, and $\upomega_w$ is the word vector of word $w$. 

For each word $w$ of document $d$, a binary indicator variable $\mathbb{S}_{d,w}$ is sampled from a Bernoulli distribution to determine whether Dirichlet multinomial model or latent feature model will be used to generate $w$. The generative process is described as follows:

(1). Sample a topic proportion $\theta\sim Dirichlet(\alpha)$.

(2). For each topic $k\in\left\{1,...,K\right\}$:

    \qquad (a) Draw a topic-word distribution $\theta_k\sim Dirichlet(\beta)$.

(3). For each document $d \in \bm{D}$:

    \qquad (a) Sample a topic $z_d \sim Multinomial(\theta)$.

    \qquad (b) For each word $w\in\left\{w_{d,1},...,w_{d,n_d}\right\}$:
           
           \qquad\qquad (i) Sample a binary indicator variable $\mathbb{S}_{d,w} \in Bernoulli(\lambda)$

           \qquad\qquad (ii) Sample a word $w \sim (1-s_w)Multinomial(\phi_{z_d}) + s_w(\sigma(\uptau_{z_d}\bm{\upomega}^T))$.

Here, the hyper-parameter $\lambda$ is the probability of a word being generated by latent feature model, and $\mathbb{S}_{d,w}$ indicates whether Dirichlet multinomial model or latent feature model is applied to word $w$ of document $d$. The topic feature (TF) in LF-DMM is similar with GSDMM, so we do not present the addible and deletable properties for LF-DMM.

Based on collapsed Gibbs sampling, LF-DMM uses the following conditional probability distribution to infer the topic of the document $d$, 
\begin{equation} \label{lfdmm_z}
\begin{split}
p&(z_d=k|\bm{Z}_{\neg d},\bm{D},\uptau,\bm{\upomega}) \propto(m_{k,\neg d}+\alpha)\\
&\prod_{w\in d}((1-\lambda)\frac{n_{k,\neg{d}}^w + \beta}{n_{k,\neg{d}}+V\beta} + \lambda \sigma(w|\uptau_k\bm{\upomega}^T))^{n_d^w}
\end{split}
\end{equation}
where $n_d^w$ is the number of word $w$ in document $d$.

The binary indicator variable $\mathbb{S}_{d,w}$ for word $w$ in document $d$ conditional on $z_d=k$ is inferred using the following distribution,
\begin{equation} \label{lfdmm_b}
p({\mathbb{S}_{d,w}=s|z_d=k)}\propto
\begin{cases}
(1-\lambda)\frac{n_{k, \neg d}^{w_i}+\beta }{{n_{k,\neg d}+V\beta}} \ \ for \  s=0,\\
\lambda \sigma(w_i|{\uptau_k}\bm{\upomega}^T) \ \ for \ s=1.
\end{cases}
\end{equation}
where the subscript $\neg{d}$ means document $d$ is removed from its current topic feature (TF) vector.

After each iteration, LF-DMM estimates the topic vectors using the following optimization function,

\begin{equation} \label{lfdmm_mle}
\begin{split}
L_k = - \sum_{w\in W} F_k^w & (\uptau_k \cdot \upomega_w - \log(\sum_{w'\in W}e^{\uptau_k\cdot\upomega_{w'}})) \\ 
& + \mu||\uptau_k||_2^2
\end{split}
\end{equation}
where $F_k^w$ is the number of times word $w$ generated from topic $k$ by latent feature model. LF-DMM adopted L-BFGS \footnote{LF-DMM used the implementation of the Mallet toolkit \cite{mccallum2002mallet}} \cite{liu1989limited} to find the topic vector $\uptau_k$ that minimizes $L_k$.

\subsubsection{GPU-DMM}

Based on DMM model, GPU-DMM \cite{li2016topic} promotes the semantically related words under the same topic during the sampling process by the generalized P\'{o}lya urn (GPU) model \cite{mahmoud2008polya}. When a ball of a particular color is sampled, a certain number of balls of similar colors are put back along with the original ball and a new ball of that color. In this case, sampling a word $w$ in topic $k$ not only increases the probability of $w$ itself under topic $k$, but also increases the probability of the semantically similar words of word $w$ under topic $k$.

Given pre-trained word embeddings, the semantic similarity between two words $w_i$ and $w_j$ is denoted by $cos(w_i,w_j)$ that are measured by cosine similarity. For all word pairs in vocabulary, if the semantic similarity score is higher that a predefined threshold $\epsilon$, the word pair is saved into a matric $\mathbb{M}$, i.e., $\mathbb{M}=\{(w_i,w_j)|cos(w_i,w_j)>\epsilon\}$. Then, the promotion matrix $\mathbb{A}$ with respect to each word pair is defined below,

\begin{equation} \label{gpudmm_pro}
\mathbb{A}_{w_i,w_j}=
\begin{cases}
  1  & \ \ \ w_i=w_j \\
  \mu   &  \ \ \ w_j \in \mathbb{M}_{w_i} \ and \ w_j \neq w_i \\
  0  & \ \ \ otherwise x
\end{cases}
\end{equation}
where $\mathbb{M}_{w_i}$ is the row in $\mathbb{M}$ corresponding to word $w_i$ and $\mu$ is the pre-defined promotion weight.

GPU-DMM and DMM share the same generative process and graphical representation but differ in the topic inference process that they use. Different from DMM and LF-DMM, GPU-DMM first samples a topic for a document, and then only reinforces only the semantically similar words if and only if a word has strong ties with the sampled topic. Therefore, a nonparametric probabilistic sampling process for word $w$ in document $d$ is as follows:

\begin{equation} \label{gpu}
\mathbb{S}_{d,w} \sim Bernoulli(\lambda_{w,z_d})
\end{equation}
\begin{equation}
\lambda_{w,z_d} = \frac{p(z|w)}{p_{max}(z'|w)}
\end{equation}
\[
p_{max}(z'|w) = \max_kp(z=k|w)
\]
\begin{equation}
p(z=k|w)=\frac{p(z=k)p(w|z=k)}{\sum_{i=1}^Kp(z=i)p(w|z=i)}
\end{equation}
where $\mathbb{S}_{d,w}$ indicates whether GPU is applied to word $w$ of document $d$ given topic $z_d$. We can see that GPU model is more likely to be applied to $w$ if word $w$ is highly relate to topic $z_d$.

The Topic feature vector of a topic $k$ in GPU-DMM is defined as a tuple $\{\widetilde{n}_k^w(w\in W), m_k, \widetilde{n}_k\}$.

TF makes the same changes with GSDMM when no GPU is applied, namely $\mathbb{S}_{d,w}=0$. Under $\mathbb{S}_{d,w}=1$, the addible and deletable properties of topic feature (TF) in GPU-DMM are described below. 

(1) \textbf{Addible Property}. A document $d$ will be added into topic $k$ by updating its TF vector as follows,

\[
\begin{split}
& \widetilde{n}_k = \widetilde{n}_k + n_d^{w_i}\cdot\mathbb{A}_{w_i,w_j} \ \ \  for \  each \ word \ w_j\in\mathbb{M}_{w_i}  \\
& \widetilde{n}_k^{w_j} = \widetilde{n}_k^{w_j} +  n_d^w\cdot\mathbb{A}_{w_i,w_j} \ \ \  for \  each \ word \ w_j\in\mathbb{M}_{w_i}  \\
& m_k = m_k + 1
\end{split}
\]

(2) \textbf{Deletable Property}. A document $d$ will be deleted from topic $k$ by updating its TF vector as follows,

\[
\begin{split}
& \widetilde{n}_k = \widetilde{n}_k - n_d^{w_i}\cdot\mathbb{A}_{w_i,w_j} \ \ \  for \  each \ word \ w_j\in\mathbb{M}_{w_i}  \\
& \widetilde{n}_k^{w_j} = \widetilde{n}_k^{w_j} - n_d^w\cdot\mathbb{A}_{w_i,w_j} \ \ \  for \  each \ word \ w_j\in\mathbb{M}_{w_i}  \\
& m_k = m_k - 1
\end{split}
\]

Accordingly, based on Gibbs sampling, the conditional distribution to infer the topic for each document in Equation \ref{equation1} is rewritten as follows:

\begin{equation}
\begin{split}
 p(z_d=k|&\bm{Z}_{\neg d},\bm{D})\propto \frac{m_{k,\neg{d}}+\alpha}{N-1+K\alpha} \times\\
 & \frac{\prod_{w\in d}\prod_{j=1}^{n_d^w}(\widetilde{n}_{k,\neg{d}}^w+\beta+j-1)}{\prod_{i=1}^{n_{d}}(\widetilde{n}_{k,\neg{d}}+V\beta+i-1)}\\
\end{split}
\end{equation}

During each iteration, GPU-PDMM first delete it from its current TF vector with the deletable property. After obtaining the topic of the document, GPU-DMM first updates $\mathbb{S}_{d,w}$ for GPU using Equation $\ref{gpu}$, and then updates TF vector for each word using the addible property. Finally, the posterior distribution in Equation \ref{equation2} for GPU-DMM is rewritten as follows:

\begin{equation} \label{gpudmm_w}
\phi_k^w=\frac{\widetilde{n}^w_k+\beta}{\widetilde{n}_k+V\beta}
\end{equation}

\subsubsection{GPU-PDMM}

Considering the single-topic assumption may be too strong for some short text corpus, Li et al. \cite{li2017enhancing} first proposed Poisson-based Dirichlet Multinomial Mixture model (PDMM) that allows each document can be generated by one or more (but not too many) topics. Then PDMM can be extended as GPU-PDMM model by incorporating generalized P\'{o}lya urn (GPU) model during the sampling process. 

\begin{figure}[H]
  \centering
  \includegraphics[width=80mm]{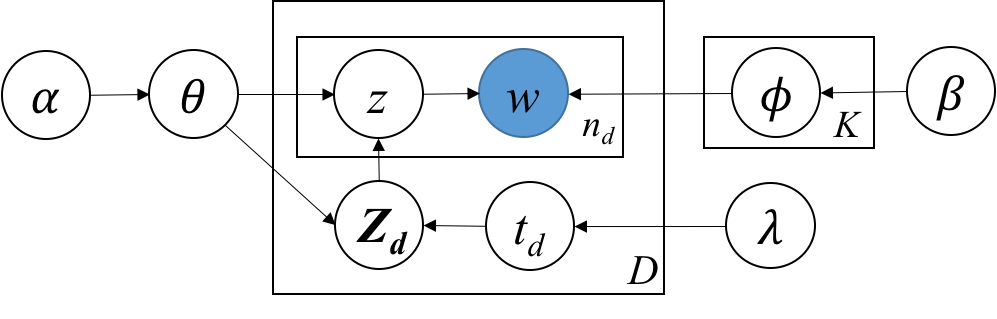}
  \caption{Graphical Model of GPU-PDMM.}\label{pdmm}
\end{figure}

In GPU-PDMM, each document is generated by $t_d$ ($0<t_d\leq \varsigma$) topics, where $\varsigma$ is the maximum number of topics allowable in a document. GPU-PDMM uses Poisson distribution to model $t_d$. The graphical model of GPU-PDMM is shown in Figure \ref{pdmm}. The generative process of GPU-PDMM is described as follows. 

(1). Sample a topic proportion $\theta\sim Dirichlet(\alpha)$.

(2). For each topic $k\in\left\{1,...,K\right\}$:

    \qquad (a) Draw a topic-word distribution $\theta_k\sim Dirichlet(\beta)$.

(3). For each document $d \in \bm{D}$:

    \qquad (a) Sample a topic number $t_d\sim Poisson(\lambda)$.

    \qquad (b) Sample $t_d$ distinct topics $\textbf{Z}_d\sim Multinomial(\theta)$.

    \qquad (c) For each word $w\in\left\{w_{d,1},...,w_{d,n_d}\right\}$:

           \qquad\qquad (i) Uniformly sample a topic $z_{d,w}\sim \textbf{Z}_d$.

           \qquad\qquad (ii) Sample a word $w\sim Multinomial(\phi_{z_{d,w}})$.

Here $t_d$ is sampled using Poisson distribution with parameter $\lambda$, and $\textbf{Z}_d$ is the topic set for document $d$.

The topic feature (TF) vector of a topic $k$ in GPU-PDMM is defined as a tuple {$\widetilde{n}_k, \widetilde{n}_k^w(w\in W), c_k, d_k, n_{k,d}, n_{k,d}^w\}$, where $c_k$ is the number of words associated with topic $k$ and $d_k$ represents the word set in topic $k$.

The addible and deletable properties of topic feature (TF) in GPU-PDMM are described below. The $\widetilde{n}_k$ and $\widetilde{n}_k^w$ of TF in GPU-PDMM makes the same changes with GPU-DMM. Here, we only describe other variables in TF.

(1) \textbf{Addible Property}. Suppose that word $w$ in document $d$ will be added to topic $k$, TF feature is updates as follows,

\[
\begin{split}
 & c_k = c_k + 1 \ ; \ d_k = d_k + {w} \\
 & n_{k,d} = n_{k,d} + 1 \ ; \ n_{k,d}^w = n_{k,d}^w + 1 \\
\end{split}
\]

(2) \textbf{Deletable Property}. Suppose that word $d$ will be deleted from topic $k$. TF feature is updated as follows,

\[
\begin{split}
 & c_k = c_k - 1 \ ; \ d_k = d_k - {w} \\
 &  n_{k,d} = n_{k,d} - 1 \ ; \ n_{k,d}^w = n_{k,d}^w - 1 \\
\end{split}
\]

The Gibbs sampling process of GPU-PDMM is similar to GPU-DMM, it updates the topic for word $w$ in document $d$ using the following equation,
\begin{equation} \label{gpupdmm_w}
p(z_{d,w}=k|z_{\neg(d,w)},\textbf{Z}_d, \bm{D})\propto\frac{1}{t_d}\times\frac{\widetilde{n}_{k,\neg(d,w)}^w+\beta}{\sum_w^V\widetilde{n}_{k,\neg(d,w)}^w+V\beta}
\end{equation}

Conditioned on all $z_{d,w}$ in document $d$, GPU-PDMM samples each possible $\textbf{Z}_d$ as follows,

\begin{equation} \label{gpupdmm_d}
\begin{split}
p(\textbf{Z}_d|& \bm{Z}_{\neg{d}}, \bm{D})\propto \frac{\lambda^{t_d}}{t_d^{n_d}}\\
&\times \frac{\prod_{k\in \textbf{Z}_d}(c_{k,\neg{d}}+\alpha)}{\prod_{i=0}^{t_d-1}(\sum_k^Kc_{k,\neg{d}}+K\alpha-i)}\\
&\times \prod_{k\in \textbf{Z}_d}
\frac{\prod_{w\in d_k} \prod_{i=0}^{n_{k,d}^w}(\widetilde{n}_{k,\neg{d}} + n_{k,d}^w)-i+\beta}
{\prod_{i=0}^{n_{k,d}-1}(\sum_w^V \widetilde{n}_{k,\neg{d}}^w+n_{k,d}-i+V\beta)}. 
\end{split}
\end{equation}

During each iteration, for each document $d$, GPU-PDMM first updates TF vector using Deletable Property and the topic for each word $w$ in $d$ using Equation $\ref{gpupdmm_w}$. Then GPU-PDMM samples each possible $\textbf{Z}_d$ using Equation $\ref{gpupdmm_d}$. Finally, GPU-PDMM sets all the values of $z_{d,w}$ based on the updates $\textbf{Z}_d$, updates $\mathbb{S}_{d,w}$ for GPU using Equation $\ref{gpu}$, and then updates TF vector for each word using the addible property.

Here, due to the computational costs involved in sampling $\textbf{Z}_d$, GPU-PDMM only samples the more relevant topics for each document. Specifically, GPU-PDMM infers the topic probability $p(z|d)$ of each document $d$ using the follows,

\[
p(z=k|d) \propto \sum_{w \in d} p(z=k|w)p(w|d)
\]
where $p(w|d)=\frac{n_d^w}{n_d}$. GPU-PDMM only chooses the top $M$ topics for document $d$ based on the probability $p(z|d)$ to generate $\textbf{Z}_d$, where $\varsigma<M\leq K$. The topic-word distribution can be calculated by Equation \ref{gpudmm_w}.

\subsection{Global Word Co-occurrences based Methods}

The closer the two words, the more relevance the two words. Utilizing this idea, global word co-occurrences based methods learn the latent topics from the global word co-occurrences obtained from the original corpus. This type of methods needs to set sliding window for extracting word co-occurrences. In general, if the average length of each document is larger than 10, they use sliding window and set the size of the sliding window as 10, else they can directly take each document as a sliding window. 

\subsubsection{BTM}

BTM \cite{cheng2014btm} first generate biterms from the corpus $\bm{D}$, where any two words in a document is treated as a biterm. Suppose that the corpus contains $n_b$ biterms $\textbf{B}=\{b_i\}_{i=1}^{n_B}$, where $b_i=(w_{i,1},w_{i,2})$. BTM infers topics over the biterms $B$. The generative process of BTM is described as follows, and its graphical model is shown in Figure \ref{btm}.

\begin{figure}[H]
  \centering
  \includegraphics[width=80mm]{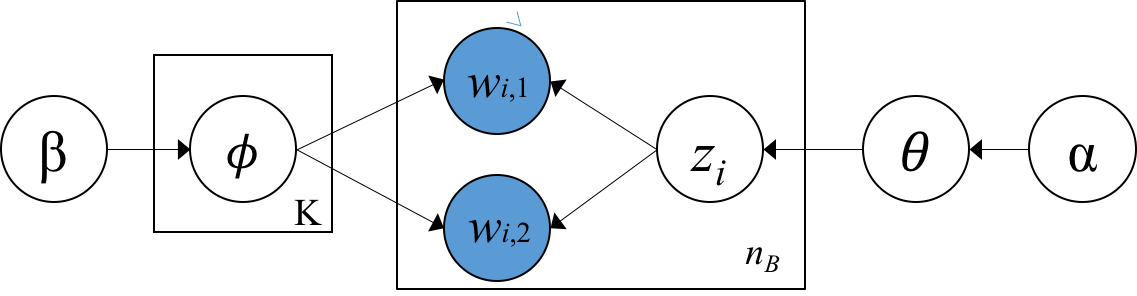}
  \caption{Graphical Model of BTM.} \label{btm}
\end{figure}

(1). Draw $\theta \sim$ Dirichlet($\alpha$).

(2). For each topic $k\in[1,K]$

    \qquad (a) draw $\phi_k\sim$ Dirichlet($\beta$).

(3). For each biterm $b_i\in \textbf{B}$

   \qquad (a) draw $z_i \sim$ Multinomial($\theta$),

   \qquad (b) draw $w_{i,1},w_{i,2} \sim$ Multinomial($\phi_{z_i}$).

The TF vector of a topic $k$ in BTM is defined as a tuple $\{n_k(w\in W), n_k\}$. The addible and deletable properties of topic feature (TF) in BTM are described below.

(1) \textbf{Addible Property}. A biterm $b_i$ can be efficiently added to topic $k$ by updating its TF vector as follows.

\[
\begin{split}
  n^{w_{i,1}}_k = n^{w_{i,1}}_k+ 1 \ ; \ n^{w_{i,2}}_k = n^{w_{i,2}}_k + 1 \ ; \ n_k = n_k + 1 
\end{split}
\]

(2) \textbf{Deletable Property}. A biterm $b_i$ can be efficiently deleted from topic $k$ by unpdating its TF vector as follows.

\[
\begin{split}
  n^{w_{i,1}}_k = n^{w_{i,1}}_k - 1 \ ; \ n^{w_{i,2}}_k = n^{w_{i,2}}_k - 1 \ ; \ n_k = n_k - 1
\end{split}
\]

Using the technique of collapsed Gibbs sampling, BTM samples the topic $z_i$ of biterm $b_i$ using the following conditional distribution,

\begin{equation} \label{btm_b}
\begin{split}
p(z_{i}=k|& \bm{Z}_{\neg{i}}, \textbf{B})\propto(n_{k,-i}+\alpha) \times\\
&\frac{(n_{k,\neg{i}}^{w_{i,1}}+\beta)(n_{k,\neg{i}}^{w_{i,2}}+\beta)}{(n_{k,\neg{i}}+V\beta+1)(n_{k,\neg{i}}+V\beta)}
\end{split}
\end{equation}
where $\bm{Z}_{\neg{i}}$ denotes the topics for all biterms except the current biterm $b_i$, and $n_k$ is the number of biterms assigned to topic $k$.

 For each biterm, we first delete it from its current TF vector with the deletable property. Then, we reassign the biterm to a topic using Equation \ref{btm}. Accordingly, we update the new TF vector with the addible property. After finishing the iterations, BTM estimates $\phi$ and $\theta$ as follows,

\begin{equation}
\phi_k^w=\frac{n_k^w+\beta}{n_k+V\beta},
\end{equation}
\begin{equation}
\theta_d^k=\sum_{i=1}^{n_d^b}p(z_i=k)
\end{equation}
where $\theta_d^k$ is the probability of topic $k$ in document $d$, and $n_d^b$ is the number of biterms in document $d$.

\subsubsection{WNTM}

WNTM \cite{zuo2016word} uses global word co-occurrence to construct word co-occurrence network, and learns the distribution over topics for each word from word co-occurrence network using LDA. WNTM first set the size of a sliding window, and the window is moving word by word. Suppose window size is set as 10 in the original paper, if one document has 15 words, it will have 16 windows in this document. As WNTM scanning word by word in one window, two distinct words in the window are regarded as co-occurrence. WNTM construct undirected word co-occurrence network, where each node of the word co-occurrence network represents one word and the weight of each edge is the number of co-occurrence of the two connected words. We can see that the number of nodes is the number of vocabulary $V$. 

 Then, WNTM generates one pseudo-document $l$ for each vertex $v$ which is consisted of the adjacent vertices of this vertex in word network. The occur times of this adjacent vertex in $l$ is determined by the weight of the edge. The number of words in $l$ is the degree of the vertex $v$ and the number of pseudo-documents $P$ is the number of vertices.

After obtaining pseudo-documents $P$, WNTM adopts LDA to learn latent topics from pseudo-documents. Therefore, the topic feature (TF) in LF-DMM is same with LDA. For each word $w$ in $l$, WNTM infers its topic using the following conditional distribution,

\begin{equation} 
p(z_{l,w}=k \mid \bm{Z}_{\neg{(l,w)}}, P, \alpha, \beta) \propto (n_{l,\neg{(l,w)}}^{k} + \alpha)\dfrac{ n^{w}_{k,\neg{(l,w)}}+\beta}{n_{k,\neg{(l,w)}}+V\beta} 
\end{equation}
where $n_l^k$ is the number of topic $k$ belonging to pseudo-document $l$, and $\neg{(l,w)}$ means word $w$ is removed from its pseudo-document $l$.

Because each pseudo-document is each word's adjacent word-list, the document-topic distribution learned from pseudo-document is the topic-word distribution in WNTM. Suppose pseudo-document $l$ is generated from word $w$, the topic-word distribution of $w$ is calculated using the following Equation,

\begin{equation}
\phi_k^w=\frac{n_l^k+\alpha}{n_l+K\alpha}
\end{equation}
where $n_l$ is the number of words in $l$.

Given topic-word distribution, the document-word distribution $\theta_d$ can be calculated as,

\[
\theta_d^k=\sum_{i=1}^{n_d}\phi_k^{w_{d,i}}p(w_{d,i}|d)
\]

\[
p(w_{d,i}|d)=\frac{n_d^{w_{d,i}}}{n_d}
\]
where $n_d^{w_{d,i}}$ is the number of word $w_{d,i}$ in document $d$.

\subsection{Self-aggregation based Methods}

Self-aggregation based methods alleviate the problem of sparseness by merging short texts into long pseudo-documents $P$ before inferring the latent topics \cite{jin2011transferring,qiang2017topic,mehrotra2013improving}. The previous self-aggregation based methods first merged short texts, and then applied topic models. Recently, SATM and PTM simultaneously integrate clustering and topic modeling in one iteration. In general, the number of pseudo-documents $|P|$ is significantly less than the number of short texts, namely $|P| \ll N$.

\subsubsection{SATM}

Self-aggregation based topic modeling (SATM) \cite{quan2015short} supposes that each short text is sampled from an unobserved long pseudo-document, and infers latent topics from pseudo-documents using standard topic modeling. The Gibbs sampling process in SATM can be described in two indispensable steps.

The first step calculates the probability of the occurrence of a pseudo-document $l$ in $P$ conditioned on short document $d$ in short corpus, which is estimated using the mixture of unigrams model \cite{nigam2000text},

w_{d,i}

\begin{equation}
p(l|d)=\frac{p(l)\prod_{i=1}^V(\frac{n_{l}^{w_{d,i}}}{n_{l}})^{n_{d}^{w_{d,i}}}}{\sum_{m=1}^{|P|}p(m)\prod_{i=1}^V(\frac{n_{m}^{w_{d,i}}}{n_{m}})^{n_{d}^{w_{d,i}}}}
\end{equation}
where $p(l)=\frac{n_l}{N}$ represents the probability of pseudo-document $p_l$, $n_{l}^{w_{d,i}}$ is the number of word $w_{d,i}$ in pseudo-document $p_l$, and $n_{l}$ is the number of words in $p_l$.

The second step estimates draws a pair of pseudo-document label $l_{d,w}$ and topic label $z_{d,w}$ jointly for word $w$ in document $d$, which is similar with standard topic modeling (author-topic modeling) \cite{rosen-zvi2004author}. 

The addible and deletable properties of pseudo-document and topic feature (PTF) in SATM are described below.

(1) \textbf{Addible Property}. A word $w$ can be efficiently added into pseudo-document $l$ and topic $k$ by updating its TPF vector as follows.

\[
\begin{split}
 & n^{w}_l = n^w_l + 1 \ ; \ n_l^k = n_l^k + 1 \ ; \ n_l = n_l + 1 \ \ \\
 & n^w_k = n^w_k + 1 \ ; \ n_k = n_k + 1\\
\end{split}
\]

(2) \textbf{Deletable Property}. A word $w$ can be efficiently deleted from pseudo-document $l$ and topic $k$ by updating its PTF vector as follows.

\[
\begin{split}
  & n^{w}_l = n^w_l - 1 \ ; \ n_l^k = n_l^k - 1 \ ; \ n_l = n_l - 1 \\
 & n^w_k = n^w_k - 1 \ ; \ n_k = n_k - 1\\
\end{split}
\]

The pair of pseudo-document label $l_{d,w}$ and topic label $z_{d,w}$ jointly for word $w$ in document $d$ can be calculated by, 

\begin{equation}
\begin{split}
p(l_{d,w}=l,&z_{d,w}=k|\textbf{Z}_{\neg(d,w)}, P_{\neg(d,w)})\propto \\
& p(l|d) \times\frac{n_{l,\neg(d,w)}^k+\alpha}{n_{l,\neg(d,w)}+K\alpha}\cdot\frac{n_{k,\neg(d,w)}^w+\beta}{n_{k,\neg(d,w)}+V\beta}\\
\end{split}
\end{equation}
where $n_l^k$ is the number of words in pseudo-document $l$ belonging to topic $k$.

After finishing the iterations, SATM estimates $\phi$ and $\theta$ as follows,
\begin{equation}
\phi_k^w=\frac{n_k^w+\beta}{n_k+V\beta},
\end{equation}
\begin{equation}
\theta_d^k=\prod_{i=1}^{n_d}\phi_k^{w_{d,i}}
\end{equation}

\subsubsection{PTM}

The pseudo-document-based topic modeling (PTM) \cite{zuo2016topic} supposes that each short text is sampled from one long pseudo-document $p_l$, and then infers the latent topics from long pseudo-documents $P$. A multinomial distribution $\varphi$ is used to model the distribution of short texts over pseudo-documents. The graphical model of PTM is shown in Figure \ref{ptm}. The generative process of PTM is described as follows,

\begin{figure}[H]
  \centering
  \includegraphics[width=80mm]{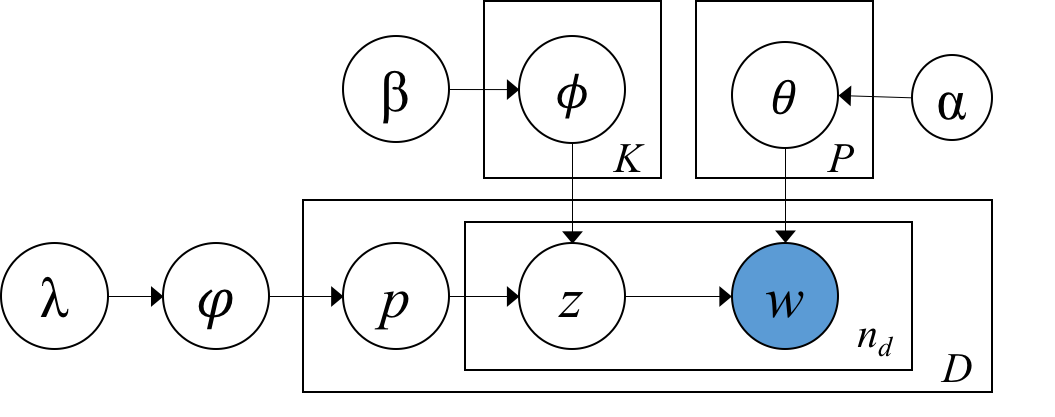}
  \caption{Graphical Model of PTM.}\label{ptm}
\end{figure}

(1). Sample $\varphi \sim Dir(\lambda)$

(2). For each topic $k\in[1,K]$

    \qquad (a) draw $\phi_k\sim$ Dirichlet($\beta$).

(3). For each pseudo-document $l$
    
    \qquad (a) sample $\theta_l \sim Dir(\alpha)$

(4). For each document $d \in \bm{D}$:

    \qquad (a) Sample a pseudo-document $l \sim Multinomial(\varphi)$.

    \qquad (b) For each word $w\in\left\{w_{d,1},...,w_{d,n_d}\right\}$ in $d$:

           \qquad \qquad (i) Sample a topic $z\sim Multinomial(\theta_{l})$.

           \qquad \qquad (ii) Sample a word $w\sim Multinomial(\phi_z)$.

The addible and deletable properties of pseudo-document and topic feature (PTF) in PTM are described below.

(1) \textbf{Addible Property}. A document $d$ can be efficiently added into pseudo-document $l$ by updating its PTF vector as follows.

\[
\begin{split}
& n^k_l = n^k_l + 1 \ \ \ for \ z_{d,w}=k \ in \ d \\
&\ m_l = m_l + 1 \ ; \ n_l = n_l + n_d\\
 \end{split}
\]

(2) \textbf{Deletable Property}. A document $d$ can be efficiently deleted from pseudo-document $l$ by updating its PF vector as follows.

\[
\begin{split}
& n^k_l = n^k_l - 1 \ \ \ for \ z_{d,w}=k \ in \ d \\
&\ m_l = m_l - 1 \ ; \ n_l = n_l - n_d\\
 \end{split}
\]

Integrating out $\theta, \phi$ and $\varphi$, the pseudo-document assignment $l$ for short text $d$ based on collapsed Gibbs sampling can be estimated as follows,

\begin{equation} \label{ptm_l}
\begin{split}
 p(l_d = l| & \overrightarrow{P_{\neg{d}}},\bm{D}) \propto \\
 &\frac{m_{l,\neg{d}}}{N-1+\lambda|P|} \frac{\prod_{k\in d}\prod_{j=1}^{n_d^k}(n_{l,\neg{d}}^k+\alpha+j-1)}{\prod_{i=1}^{n_{d}}(n_{{l},\neg{d}}+K\alpha+i-1)}\\
\end{split}
\end{equation}
where $m_l$ is the number of short texts associated with pseudo-document $l$, $n_l^k$ is the number of words associated with topic $k$ in pseudo-document $l$.

After obtaining the pseudo-document for each short text, PTM samples the topic assignment for each word $w$ in document $d$. That is,

\begin{equation} \label{ptm_z}
p(z_{d,w} = k | \textbf{Z}_{\neg(d,w)},\bm{D}) \propto (n_l^k+\alpha) \frac{n_k^{w} + \beta}{n_k + V\beta}\\
\end{equation}
where $n_{l}^k$ is the number of words associated with topic $k$ in pseudo-document $l$. 

The document-word distribution $\theta_d$ can be calculated as,
\begin{equation} \label{ptm_z}
\theta_d^k = \frac{n_{d}^k+\alpha}{n_d+K\alpha}
\end{equation}

\section{Applications}

With the emerging of social media, topic models have been used for social media content analysis, such as content characterizing and recommendation, text classification, event tracking. However, although the corpus is composed of short texts, some previous work directly applied traditional topic models for topic discovery, since no specific short text topic models were proposed at that time. Therefore, it brings a new chance for short text topic modeling to improve the performance of these tasks.

\subsection{Content characterizing and recommendation}

Microblogging sites are used as publishing platforms to create and consume content from sets of users with overlapping and disparate interests, which results in many contents are useless for users. These work \cite{ramage2010characterizing,zhao2011comparing} have been devoted to content analysis of Twitter. Ramage et al. \cite{ramage2010characterizing} used topic models to discover latent topics from the tweets that can be roughly categorized into four types: substance topics about events and ideas, social topics recognizing language used toward a social end, status topics denoting personal updates, and style topics that embody broader trends in language usage. Next, they characterize selected Twitter users along these learned dimensions for providing interpretable summaries or characterizations of users’ tweet streams. Zhao et al. \cite{zhao2011comparing} performed content analysis on tweets using topic modeling to discover the difference between Twitter and traditional medium.

Content analysis is crucial for content recommendation for microblogging users \cite{guy2015social,qian2014personalized}. Phelan et al. \cite{phelan2009using} identified emerging topics of interest from Twitter information using topic modeling, and recommended news by matching emerging topics and recent news coverage in an RSS feed. Chen et al. \cite{chen2010shorttweet} also studied content recommendation based on Twitter for better capture users' attention by exploring three separate dimensions in designing such a recommender: content sources, topic interest models for users, and social voting.

\subsection{Text Classification}

Topic models for text classification are mainly from the following two aspects. The first one is topics discovered from external large-scale data corpora are added into short text documents as external features. For example, Phan et al. \cite{phan2008learning} built a classifier on both a set of labeled training data and a set of latent topics discovered from a large-scale data collection. Chen et al. \cite{chen2011short} integrated multi-granularity hidden topics discovered from short texts and produced discriminative features for short text classification. Vo et al. \cite{vo2015learning} explored more external large-scale data collections which contain not only Wikipedia but also LNCS and DBLP for discovering latent topics. 

The other one is that topic models are used to obtain a low dimensional representation of each text, and then classify text using classification methods \cite{dai2013crest,razavi2014text}. Compared with traditional statistical methods, the representation using topic models can get a compact, dense and lower dimensional vector in which each dimension of the vector usually represents a specific semantic meaning (e.g., a topic) \cite{nigam2000text}. Dai et al. \cite{dai2013crest} used the topic information from training data to extend representation for short text. Recent topic modeling methods on text representation have explicitly evaluated their models on this task. They showed that a low dimensional representation for each text suffices to capture the semantic information.

\subsection{Event Tracking}

Nowadays, a large volume of text data is generated from the social communities, such as blogs, tweets, and comments. The important task of event tracking is to observe and track the popular events or topics that evolve over time \cite{lin2010pet,aggarwal2012event,ritter2012open}. Lin et al. \cite{lin2010pet} proposed a novel topic modeling that models the popularity of events over time, taking into consideration the burstiness of user interest, information diffusion in the network structure, and the evolution of latent topics. Lau et al. \cite{lau2012line} designed a novel topic modeling for event detecting, whose model has an in-built update mechanism based on time slices by implementing a dynamic vocabulary. 

\section{A Java Library For Short Text Topic Modeling}

We released an open-source Java library, STTM (Short Text Topic Modeling) \footnote{https://github.com/qiang2100/STTM}, which is the first comprehensive open-source library, which not only includes the state-of-the-art algorithms with a uniform easy-to-use programming interface but also includes a great number of designed modules for the evaluation and application of short text topic modeling algorithms. The design of STTM follows three basic principles. (1) Preferring integration of existing algorithms rather than implementing them. If the original implementations are open, we always attempt to integrate the original codes rather than implement them. The work that we have done is to consolidate the input/output file formats and package these different approaches into some newly designed java classes with a uniform easy-to-use member functions. (2) Including traditional topic modeling algorithms for long texts. The classical topic modeling algorithm (LDA \cite{blei2003latent} and its variation LF-LDA \cite{nguyen2015improving}) are integrated, which is easy for users to the comparison of long text topic modeling algorithms and short text topic modeling algorithms. (3) Extendibility. Because short text topic modeling is an emerging research field, many topics have not been studied yet. For incorporating future work easily, we try to make the class structures as extendable as possible when designing the core modules of STTM.

\begin{figure*}
    \centering 
    \includegraphics[scale=0.6]{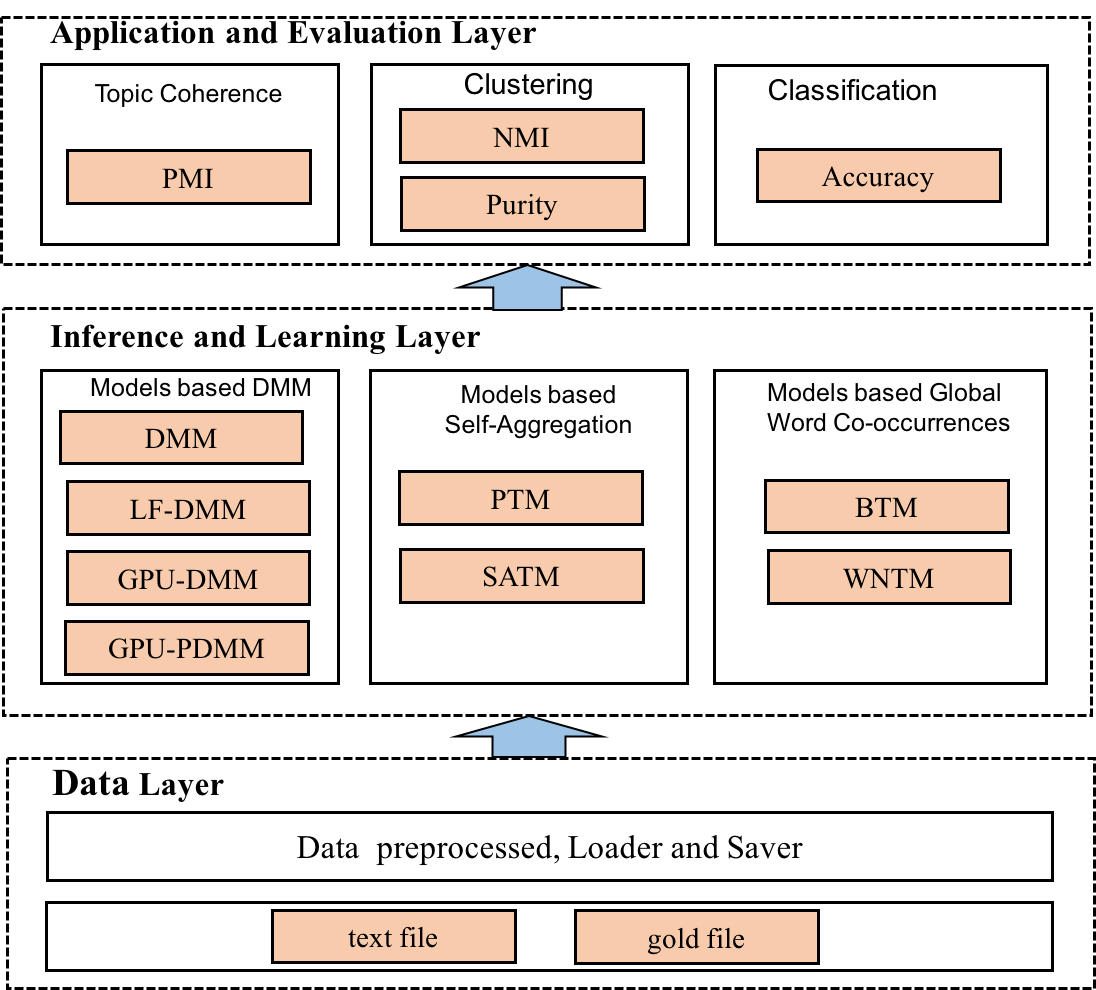}
    \caption{The architecture of STTM}\label{frame}
\end{figure*}

Figure \ref{frame} shows the hierarchical architecture of STTM. STTM supports the entire knowledge discovery procedure including analysis, inference, evaluation, application for classification and clustering. In the data layer, STTM is able to read a text file, in which each line represents one document. Here, a document is a sequence of words/tokens separated by whitespace characters. If we need to evaluate the algorithm, we also need to read a gold file, in which each line is the class label of one document. STTM provides implementations of DMM\cite{yin2014dirichlet}, LF-DMM \cite{nguyen2015improving}, GPU-DMM\cite{li2016topic}, GPU-PDMM \cite{li2017enhancing}, BTM\cite{cheng2014btm}, WNTM\cite{zuo2016word}, SATM \cite{quan2015short}, and PTM \cite{zuo2016topic}. For each model, we not only provide how to train a model on existing corpus but also give how to infer topics on a new/unseen corpus using a pre-trained topic model. In addition, STTM presents three aspects of how to evaluate the performance of the algorithms (i.e., topic coherence, clustering, and classification). For topic coherence, we use the point-wise mutual information (PMI) to measure the coherence of topics \cite{zuo2016topic}. Short text topic modeling algorithms are widely used for clustering and classification. In the clustering module, STTM provides two measures (NMI and Purity) \cite{yan2015probabilistic}. Based on the latent semantic representations learned by short text topic modeling, accuracy metric is chosen in classifications \cite{cheng2014btm}.

\section{Experimental Setup}
In this section, we specify the parameter setting of the introduced short text topic models, dataset and evaluation metrics we used. All of these are implemented in our library STTM. The experiments were performed on a Ubuntu 18.04(bionic) system with 6 cores, Intel Xeon E5645 CPU and 12288 KB cache.

For all models in comparison, we used the recommended setting by the authors and set the number of iterations as 2,000 unless explicitly specified elsewhere. The word embeddings of LF-DMM, GPU-DMM, and GPU-PDMM are trained by Glove \cite{pennington2014glove}. In this paper, we used the pre-trained word embeddings "glove.6B.200d.txt", where the dimension of the vector is 200.

\subsection{Parameter Setting}

\textbf{LDA}: LDA is the most popular and classic topic modeling. We choose it as a baseline to the comparison. The hyper-parameters of LDA are set as $\alpha=0.05$ and $\beta=0.01$ that are proved in the paper (BTM). The authors tuned parameters via grid search on the smallest collection to get the best performance.

\textbf{GSDMM}: We set $k=300$, $\alpha=0.1$ and $\beta=0.1$ declared in the paper(GSDMM).

\textbf{LF-DMM}: We set $\lambda=0.6$, $\alpha=0.1$ and $\beta=0.01$ shown in their paper. We set the iterations for baseline models as 1,500 and ran the further iterations 500 times.

\textbf{GPU-DMM}: We use the hyper-parameter settings provided by the authors, $\alpha=50/k$, $\beta=0.01$. The number of iterations is 1,000 in the paper.

\textbf{GPU-PDMM}: All the settings are same as the model \textbf{GPU-DMM}. We set $\lambda=1.5$, $\varsigma=1.5$, and $M$=10 declared in the original paper.

\textbf{BTM}: The parameters $\alpha =50/K$ and $\beta =0.01$ are used, the model gets optimal performance. Each document is treated as one window.

\textbf{WNTM}: We set $\alpha=0.1$ and $\beta=0.1$ used in the original paper. The window size is set as 10 words. 

\textbf{SATM}: We set the number of pseudo-numbers as 300, and the hyper-parameters $\alpha=50/k$, $\beta=0.1$. The number of iterations is set as 1,000.

\textbf{PTM}: The hyper-parameters are $\alpha=0.1$ and $\beta=0.01$. We also set the number of pseudo-document as 1000.

\subsection{Datasets}

To show the effects and differences of the above nine models, we select the following six datasets to verify the models. After preprocessing these datasets, we present the key information of the datasets that are summarized in Table \ref{corpus}, where $K$ corresponds to the number of topics per dataset, $N$ represents the number of documents in each dataset, $Len$ shows the average length and maximum length of each document, and $V$ indicates the size of the vocabulary.

\begin{table}[H]
  \centering
  \caption{The basic information of the corpus} \label{corpus}
  \begin{tabular}{|l|l|l|l|l|}
    \hline
    $\bm{Dataset}$&{$\bm{K}$}&{$\bm{N}$}&{$\bm{Len}$}&{$\bm{V}$}\\
    \hline
    {SearchSnippets}&{8}&{12,295}&{14.4/37}&{5,547}\\
    \hline
    {StackOverflow}&{20}&{16,407}&{5.03/17}&{2,638}\\
    \hline
    {Biomedicine}&{20}&{19,448}&{7.44/28}&{4498}\\
    \hline
    {Tweet}&{89}&{2,472}&{8.55/20}&{5,096}\\
    \hline
    {GoogleNews}&{152}&{11,109}&{6.23/14}&{8,110}\\
    \hline
    {PascalFlickr}&{20}&{4,834}&{5.37/19}&{3,431}\\
    \hline
    \end{tabular}
\end{table}

 \textbf{SearchSnippets:} Given the predefined phrases of 8 different domains, this dataset was chosen from the results of web search transaction. The 8 domains are Business, Computers, Culture-Arts, Education-Science, Engineering, Health, Politics-Society, and Sports, respectively.

 \textbf{StackOverflow:} The dataset is released on Kaggle.com. The raw dataset contains 3,370,528 samples from July 31st, 2012 to August 14, 2012. Here, the dataset randomly selects 20,000 question titles from 20 different tags.

 \textbf{Biomedicine:} Biomedicine makes use of the challenge data delivered on BioASQ's official website.

 \textbf{Tweet:} In the 2011 and 2012 microblog tracks at Text REtrieval Conference (TREC), there are 109 queries for using. After removing the queries with none highly-relevant tweets, Tweet dataset includes 89 clusters and totally 2,472 tweets.

 \textbf{GoogleNews:} In the Google news site, the news articles are divided into clusters automatically. GoolgeNews dataset is downloaded from Google news site on November 27, 2013, and crawled the titles and snippets of 11,109 news articles belonging to 152 clusters.

 \textbf{PascalFlickr:} PascalFlickr dataset are a set of captions \cite{rashtchian2010collecting}, which is used as evaluation for short text clustering \cite{finegan2016effects}.

\subsection{Evaluation Metrics}

It is still an open problem about how to evaluate short text topic models. A lot of metrics have been proposed for measuring the coherence of topics in texts \cite{newman2010automatic,mimno2011optimizing}. Although some metrics tend to be reasonable for long texts, they can be problematic for short texts \cite{quan2015short}. Most conventional metrics (e.g., perplexity) try to estimate the likelihood of held-out testing data based on parameters inferred from training data. However, this likelihood is not necessarily a good indicator of the quality of the extracted topics \cite{chang2009reading}. To provide a good evaluation, we evaluate all models from many aspects using different metrics, 

\textbf{Classification Evaluation}: Each document can bed represented using document-topic distribution $p(z|d)$. Therefore, we can evaluate the performance of topic modeling using text classification. Here, we choose accuracy as a metric for classification. Higher accuracy means the learned topics are more discriminative and representative. We use a linear kernel Support Vector Machine (SVM) classifier in LIBLINEAR \footnote{https://liblinear.bwaldvogel.de/} with the default parameter settings. The accuracy of classification is computed through fivefold cross-validation on all datasets.

\textbf{Cluster Evaluation (Purity and NMI)}: By choosing the maximum of topic probability for each document, we can get the cluster label for each text. Then, we can compare the cluster label and the golden label using metric Purity and NMI \cite{huang2013dirichlet,yin2014dirichlet}.

\textbf{Topic Coherence}: Computing topic coherence, additional dataset (Wikipedia) as a single meta-document is needed to score word pairs using term co-occurrence in the paper (Automatic Evaluation of Topic Coherence). Here, we calculate the point-wise mutual information (PMI) of each word pair, estimated from the entire corpus of over one million English Wikipedia articles \cite{li2017enhancing}. Using a sliding window of 10 words to identify co-occurrence, we computed the PMI of all a given word pair. The Wikipedia corpus can be downloaded here \footnote{https://dumps.wikimedia.org/enwiki/}. Then, we can transfer the dataset from HTML to text using the code in the STTM package. Finally, due to the large size, we only choose 1,000,000 sentences from it.

\section{Experiments and Analysis}

In this section, we conduct experiments to evaluate the performance of the nine models. We run each model 20 times on each dataset and report the mean and standard deviation.

\subsection{Classification Accuracy}

Classification accuracy is used to evaluate document-topic distribution. We represent each document with its document-topic distribution and employ text classification method to assess. For DMM based methods, we use $p(z_d=k)$ to represent each document. For other models, we adopt the giving equation $\theta_d^k$. 

The classification accuracy on six datasets using nine models is shown in Figure \ref{classification}. We observe that although the performance of methods is dataset dependent, DMM based methods which utilize word embeddings outperform others, especially on Tweet and GoogleNews datasets. This is because GoogleNews and Tweet are general (not domain-specific) datasets and word embeddings used in this paper are trained in general datasets. If we try to use these models (LF-DMM, GPU-DMM, and GPU-PDMM) on domain-specific datasets, we can further improve the performance by re-training word embeddings on domain-specific datasets. 

We also observe that self-aggregation based methods are unable to achieve high accuracy, especially the SATM method. The performance of self-aggregation based methods is affected by generating long pseudo-documents. Without any auxiliary information or metadata, the error of this step of generating pseudo-documents will be amplified in the next step. 

In conclusion, these models based on the simple assumption (BTM and GSDMM) always outperform than LDA in all datasets, which indicate that two words or all words in one document are very likely to from one topic. Here we can see that the performance of other models (LF-DMM, GPU-DMM, GPU-PDMM, WNTM) are highly data set dependent. For example, WNTM achieves good performance on Tweet, GoogleNews and StackOverflow, but performs poorly on other data sets. GPU-PDMM achieves the best performance
on all data sets, except SearchSnippets. 

\begin{figure*} 
\begin{minipage}{0.3\linewidth}
\centerline{\includegraphics[width=6cm]{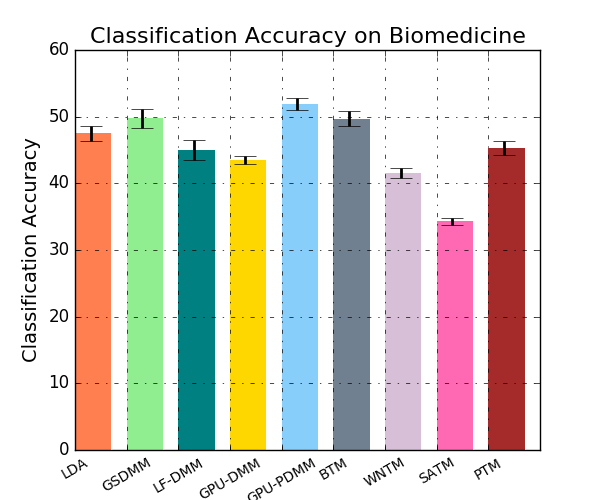}}
\centerline{(a). Biomedicine}
\end{minipage}
\hfill
\begin{minipage}{0.3\linewidth}
\centerline{\includegraphics[width=6cm]{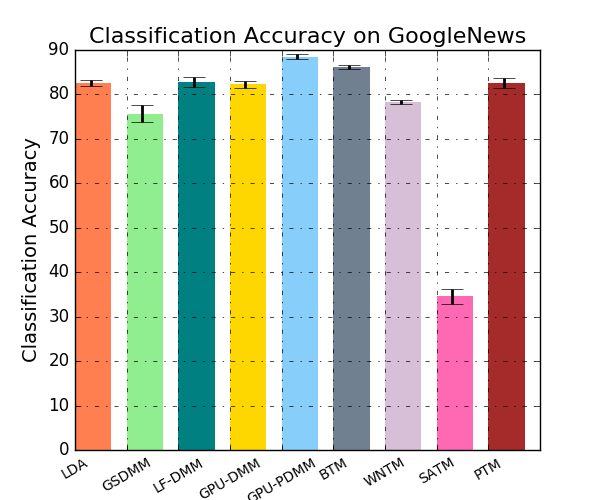}}
\centerline{(b). GoogleNews}
\end{minipage}
\hfill\begin{minipage}{0.3\linewidth}
\centerline{\includegraphics[width=6cm]{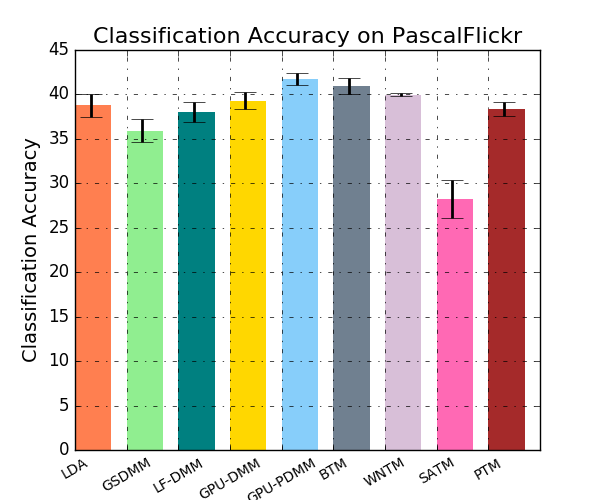}}
\centerline{(c). PascalFlickr}
\end{minipage}
\vfill
\begin{minipage}{0.3\linewidth}
\centerline{\includegraphics[width=6cm]{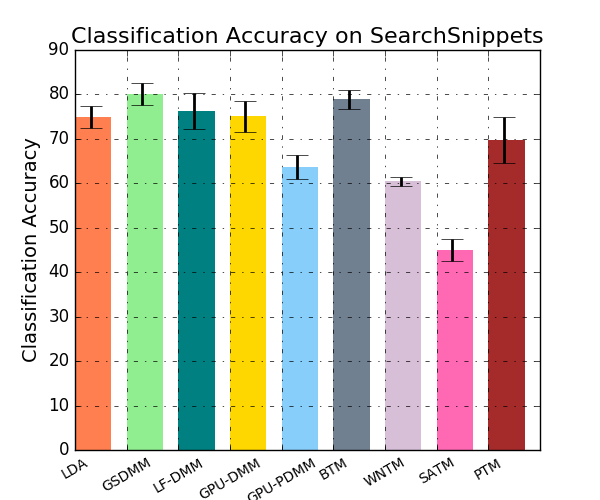}}
\centerline{(d). SearchSnippets}
\end{minipage}
\hfill
\begin{minipage}{0.3\linewidth}
\centerline{\includegraphics[width=6cm]{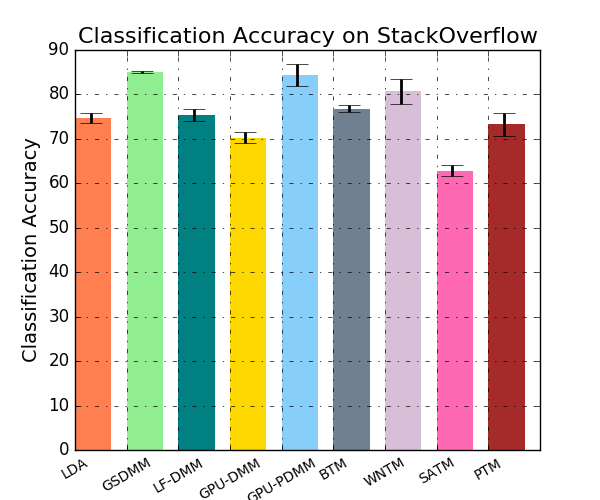}}
\centerline{(e). StackOverflow}
\end{minipage}
\hfill
\begin{minipage}{0.3\linewidth}
\centerline{\includegraphics[width=6cm]{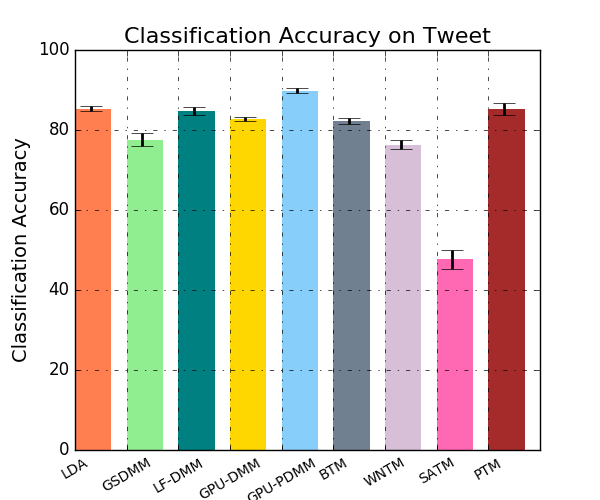}}
\centerline{(f). Tweet}
\end{minipage}
\caption{Average classification accuracy of all models on six datasets.} \label{classification}

\end{figure*}

\subsection{Clustering}

Another important application of short text topic modeling is short text clustering. For each document, we choose the maximum value from its topic distribution as the cluster label. We report the mean value of each modeling on all datasets in the last column. The best results for each dataset using each metric are highlighted in bold.

Table \ref{NMI} illustrates the results using cluster metrics. We can see that all models outperform long text topic modeling (LDA), except SATM. Here, similar to the conclusions in classification, we can observe that the performance of approaches is highly data set dependent. WNTM achieves the best performance on several datasets but performs poorly on PascalFlickr. GPU-PDMM performs very well on all datasets except SearchSnippets. 

For self-aggregation based methods, PTM performs better than SATM. For global word-occurrences based methods, two methods are very different from each other. WNTM performs better than BTM on Tweet and StackOverflow, and BTM achieves good performance on GoogleNews and PascalFlickr. For DMM based methods, GSDMM without incorporating word embeddings outperforms other methods on Biomedicine and SearchSnippets.

\begin{table*}
\centering
\caption{$Purity$ and $NMI$ value of all models on six datasets.} \label{NMI}
\begin{tabular}{|p{1.7cm}|p{1.1cm}|p{1.1cm}|p{1.1cm}|p{1.1cm}|p{1.1cm}|p{1.1cm}|p{1.1cm}|p{1.1cm}|}
\hline
\multicolumn{2}{|c|}{Model}&{Biome dicine}&{Google News}&{Pascal Flickr}
&{Search Snippets}&{Stack Overflow}
&{Tweet}&{Mean Value}
\\
\hline
\multirow{4}*{LDA}&\multirow{2}*{$Purity$}&$0.456\pm0.011$&$0.793\pm0.005$&$0.376\pm0.013$&$0.740\pm0.029$&$0.562\pm0.013$&$0.821\pm0.006$&$0.625\pm0.013$\\
\cline{2-9}
&\multirow{2}*{$NMI$}&$0.356\pm0.004$&$0.825\pm0.002$&$0.321\pm0.006$&$0.517\pm0.025$&$0.425\pm0.006$&$0.805\pm0.004$&$0.542\pm0.008$\\
\hline
\multirow{4}*{GSDMM}&\multirow{2}*{$Purity$}&$\bm{0.494\pm0.011}$&$0.754\pm0.014$&$0.360\pm0.012$&$\bm{0.801\pm0.024}$&$0.713\pm0.002$&$0.785\pm0.011$&$0.650\pm0.013$\\
\cline{2-9}
&{$NMI$}&$\bm{0.396\pm0.006}$&$0.851\pm0.004$&$0.317\pm0.005$&$\bm{0.608\pm0.023}$&$0.593\pm0.002$&$0.801\pm0.007$&$\bm{0.590\pm0.001}$\\
\hline
\multirow{4}*{LF-DMM}&\multirow{2}*{$Purity$}&$0.421\pm0.019$&$0.828\pm0.009$&$0.381\pm0.009$&$0.762\pm0.042$&$0.518\pm0.0217$&$0.856\pm0.009$&$0.630\pm0.018$\\
\cline{2-9}
&\multirow{2}*{$NMI$}&$0.348\pm0.005$&$0.875\pm0.005$&$0.365\pm0.007$&$0.579\pm0.026$&$0.443\pm0.007$&$0.843\pm0.006$&$0.578\pm0.009$\\
\hline
\multirow{4}*{GPU-DMM}&\multirow{2}*{$Purity$}&$0.433\pm0.008$&$0.818\pm0.005$&$\bm{0.395\pm0.010}$&$0.751\pm0.035$&$0.511\pm0.013$&$0.830\pm0.006$&$0.623\pm0.013$\\
\cline{2-9}
&\multirow{2}*{$NMI$}&$0.366\pm0.006$&$0.852\pm0.002$&$\bm{0.370\pm0.004}$&$0.561\pm0.026$&$0.429\pm0.003$&$0.810\pm0.006$&$0.565\pm0.008$\\
\hline
\multirow{4}*{GPU-PDMM}&\multirow{2}*{$Purity$}&$0.481\pm0.011$&$\bm{0.860\pm0.002}$&$0.368\pm0.008$&$0.537\pm0.030$&$0.702\pm0.032$&$\bm{0.869\pm0.005}$&$0.636\pm0.015$\\
\cline{2-9}
&\multirow{2}*{$NMI$}&$0.381\pm0.005$&$0.871\pm0.001$&$0.322\pm0.003$&$0.341\pm0.014$&$0.607\pm0.013$&$0.830\pm0.003$&$0.559\pm0.007$\\
\hline

\multirow{4}*{BTM}&\multirow{2}*{$Purity$}&$0.458\pm0.012$&$0.849\pm0.005$&$0.392\pm0.011$&$0.765\pm0.032$&$0.537\pm0.019$&$0.814\pm0.008$&$0.636\pm0.014$\\
\cline{2-9}
&\multirow{2}*{$NMI$}&$0.380\pm0.004$&$0.875\pm0.003$&$0.368\pm0.006$&$0.566\pm0.027$&$0.456\pm0.008$&$0.808\pm0.005$&$0.575\pm0.009$\\
\hline
\multirow{4}*{WNTM}&\multirow{2}*{$Purity$}&$0.472\pm0.009$&$0.837\pm0.007$&$0.324\pm0.005$&$0.712\pm0.016$&$\bm{0.750\pm0.026}$&$0.856\pm0.012$&$\bm{0.658\pm0.013}$\\
\cline{2-9}
&\multirow{2}*{$NMI$}&$0.369\pm0.004$&$\bm{0.876\pm0.004}$&$0.295\pm0.003$&$0.464\pm0.011$&$\bm{0.659\pm0.006}$&$\bm{0.850\pm0.009}$&$0.585\pm0.006$\\
\hline

\multirow{4}*{SATM}&\multirow{2}*{$Purity$}&$0.384\pm0.007$&$0.654\pm0.008$&$0.237\pm0.059$&$0.459\pm0.055$&$0.505\pm0.019$&$0.392\pm0.011$&$0.438\pm0.027$\\
\cline{2-9}
&\multirow{2}*{$NMI$}&$0.27\pm0.001$&$0.76\pm0.005$&$0.186\pm0.049$&$0.205\pm0.036$&$0.366\pm0.011$&$0.507\pm0.006$&$0.382\pm0.018$\\

\hline
\multirow{4}*{PTM}&\multirow{2}*{$Purity$}&$0.425\pm0.012$&$0.807\pm0.010$&$0.359\pm0.012$&$0.674\pm0.057$&$0.481\pm0.034$&$0.839\pm0.007$&$0.597\pm0.022$\\
\cline{2-9}
&\multirow{2}*{$NMI$}&$0.353\pm0.003$&$0.866\pm0.005$&$0.336\pm0.010$&$0.457\pm0.045$&$0.442\pm0.016$&$0.846\pm0.006$&$0.550\pm0.014$\\
\hline
\end{tabular}
\end{table*}

\subsection{Topic Coherence}

Topic coherence is used to evaluate the quality of topic-word distribution. Here, we only choose the top 10 words for each topic based on the word probability. The results are shown in Figure \ref{coherence}. DMM based methods achieve the best performance on all datasets. LF-DMM has the best performance on four datasets (Biomedicine, GoogleNews, SearchSnippets, and Tweet), GPU-DMM has the best performance on StackOverflow, and GPU-PDMM achieves the best on PascalFlickr. It means that incorporating word embeddings into DMM can help to alleviate the sparseness. Two methods based on global word co-occurrences perform very well and achieve a similar result on each dataset, which indicates that the adequacy of global word co-occurrences can mitigate the sparsity of short texts. Similar to the above results using other metrics, self-aggregation based methods perform very poorly. 

\begin{figure*}
\begin{minipage}{0.3\linewidth}
\centerline{\includegraphics[width=6cm]{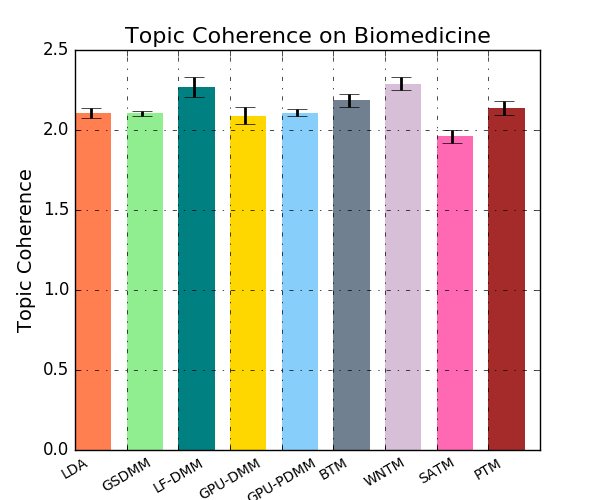}}
\centerline{(a). Biomedicine}
\end{minipage}
\hfill
\begin{minipage}{.3\linewidth}
\centerline{\includegraphics[width=6cm]{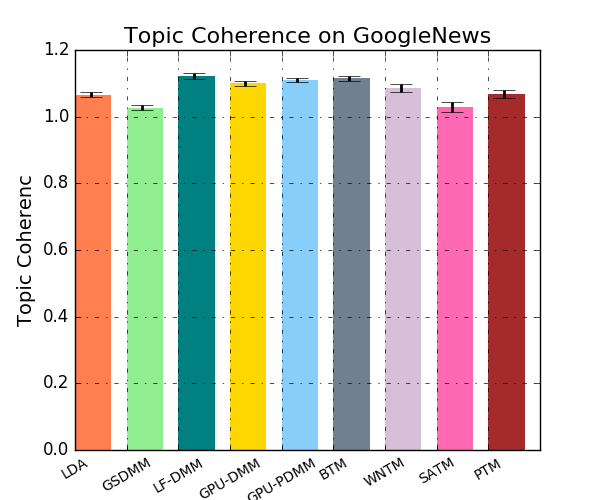}}
\centerline{(b). GoogleNews}
\end{minipage}
\hfill\begin{minipage}{0.3\linewidth}
\centerline{\includegraphics[width=6cm]{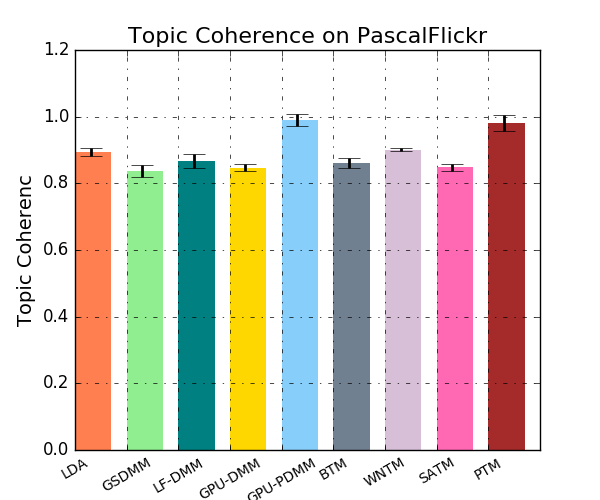}}
\centerline{(c). PascalFlickr}
\end{minipage}
\vfill
\begin{minipage}{.3\linewidth}
\centerline{\includegraphics[width=6cm]{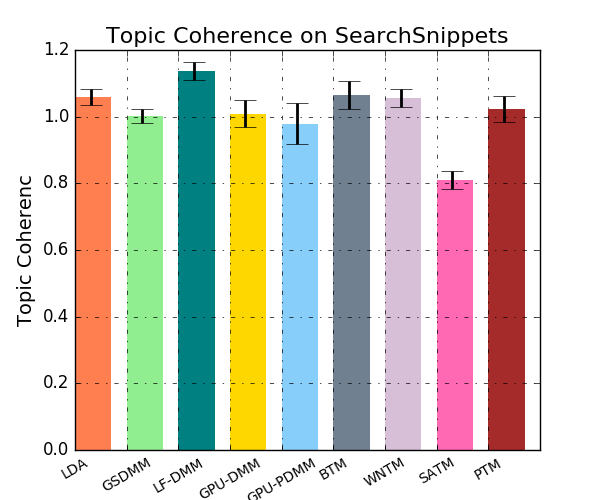}}
\centerline{(d). SearchSnippets}
\end{minipage}
\hfill
\begin{minipage}{0.3\linewidth}
\centerline{\includegraphics[width=6cm]{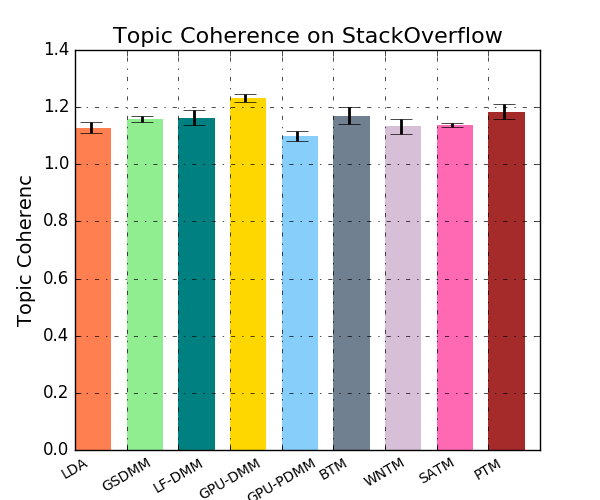}}
\centerline{(e). StackOverflow}
\end{minipage}
\hfill
\begin{minipage}{0.3\linewidth}
\centerline{\includegraphics[width=6cm]{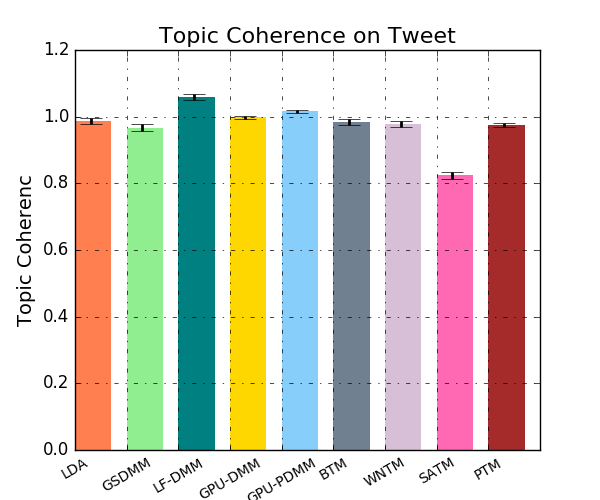}}
\centerline{(f). Tweet}
\end{minipage}
\caption{ Topic Coherence of all models on six datasets.} \label{coherence}

\end{figure*}

We also present the qualitative evaluation of latent topics. Here, we choose SearchSnippets dataset as an example, since it only contains eight topics that are Health, Politics-Society (politics), Engineering (engine.), Culture-Arts (culture), Sports, Computers, Business, and Education-Science (education). Table \ref{top10} shows the eight topics learned by the nine models. Each topic is visualized by the top ten words. Words that are noisy and lack of representativeness are
highlighted in bold. 

From Table \ref{top10}, we observe that LF-DMM can achieve a similar conclusion with topic coherence, which can learn more coherent topics with fewer noisy and meaningless words. GPU-DMM and GPU-PDMM can not discriminate the topic 'Engineering'. SATM remains the worst method in all short text topic models, which cannot discriminate three topics 'Engineering', 'Politics-Society' can 'Culture. Except LDA, PTM, WNTM, and SATM, other models can identify at least seven topics from all eight topics.

\begin{table*}
  \centering
  \caption{The top ten words of each topic by each model on SearchSnippets dataset. } \label{top10}
  \setlength{\tabcolsep}{0.5mm}{
  \begin{tabular}{|cccc|cccc|cccc|}
    \hline
    \multicolumn{4}{|c|}{LDA}&\multicolumn{4}{|c|}{GSDMM}&\multicolumn{4}{|c|}{LF-DMM}\\
    \hline
    {Topic1}&{Topic2}&{Topic3}&{Topic4}&{Topic1}&{Topic2}&{Topic3}&{Topic4}&{Topic1}&{Topic2}&{Topic3}&{Topic4}\\
    {(health)}&{(politics)}&{(engine.)}&{(culture)}&{(health)}&{(politics)}&{(engine.)}&{(culture)}&{(health)}&{(politics)}&{(engine.)}&{(culture)}\\
    \hline
    {health}&{\textbf{wikipedia}}&{car}&{music}&{health}&{political}&{car}&{movie}&{health}&{culture}&{motor}&{film}\\
    \textbf{information}&{\textbf{encyclopedia}}&{engine}&{movie}&\textbf{information}&{culture}&{engine}&\textbf{com}&{cancer}&{party}&{engine}&{music}\\
    {cancer}&{\textbf{wiki}}&{electrical}&\textbf{com}&{cancer}&{culture}&\textbf{com}&{art}&{disease}&{democratic}&{wheels}&{art}\\
    {\textbf{gov}}&\textbf{political}&\textbf{com}&\textbf{news}&{gov}&{democracy}&{electrical}&{film}&{healthy}&{war}&{electronics}&\textbf{com}\\
    {medical}&\textbf{culture}&{products}&{film}&{medical}&{\textbf{wikipedia}}&\textbf{wikipedia}&{fashion}&{medical}&{political}&{electric}&{fashion}\\
    \textbf{news}&\textbf{democracy}&{digital}&{movies}&\textbf{news}&{party}&{system}&{\textbf{motor}}&{drug}&{democracy}&{cars}&{movie}\\
    {research}&{\textbf{system}}&{home}&{yahoo}&{healthy}&{war}&{wheels}&{\textbf{wikipedia}}&{treatment}&{congress}&{models}&{books}\\
    {disease}&\textbf{party}&{motor}&{art}&{disease}&{republic}&{\textbf{olympic}}&{books}&{physical}&{presidential}&{\textbf{phone}}&{arts}\\
    {healthy}&\textbf{republic}&{energy}&{video}&{nutrition}&\textbf{information}&{digital}&{arts}&{food}&{communist}&{\textbf{graduation}}&{rock}\\
    {nutrition}&\textbf{philosophy}&{calorie}&{arts}&{hiv}&{government}&\textbf{trade}&{movies}&{care}&{philosophy}&{\textbf{fashion}}&{band}\\
    \hline
    {Topic5}&{Topic6}&{Topic7}&{Topic8}&{Topic5}&{Topic6}&{Topic7}&{Topic8}&{Topic5}&{Topic6}&{Topic7}&{Topic8}\\
    {(sport)}&{(computer)}&{(business)}&{(education)}&{(sport)}&{(computer)}&{(business)}&{(education)}&{(sport)}&{(computer)}&{(business)}&{(education)}\\
    \hline
    {com}&{computer}&\textbf{gov}&{edu}&\textbf{news}&{computer}&{business}&{research}&{sports}&{computer}&{business}&{research}\\
    \textbf{news}&{business}&\textbf{business}&{research}&{\textbf{music}}&{software}&{market}&{edu}&{football}&{intel}&{financial}&{edu}\\
    {sports}&{web}&\textbf{information}&{science}&\textbf{com}&{web}&\textbf{news}&{science}&{games}&{software}&{bank}&{graduate}\\
    {football}&{software}&{\textbf{school}}&{theory}&{sports}&{programming}&\textbf{information}&{theory}&\textbf{news}&{device}&{economic}&{resources}\\
    {games}&\textbf{com}&\textbf{trade}&{\textbf{journal}}&{football}&\textbf{wikipedia}&{stock}&\textbf{information}&{league}&{linux}&{trade}&{science}\\
    {\textbf{amazon}}&\textbf{news}&{\textbf{edu}}&{theoretical}&{games}&{memory}&{gov}&{school}&\textbf{com}&{digital}&\textbf{news}&{university}\\
    {game}&{\textbf{market}}&\textbf{research}&{physics}&{\textbf{movie}}&\textbf{com}&\textbf{com}&{university}&{hockey}&{network}&{market}&{school}\\
    {soccer}&{\textbf{stock}}&\textbf{home}&{computer}&{game}&{intel}&{finance}&{journal}&{game}&{hardware}&{services}&{faculty}\\
    {world}&{internet}&\textbf{law}&\textbf{information}&{\textbf{wikipedia}}&{internet}&{services}&{physics}&{soccer}&{web}&{law}&{center}\\
    {tennis}&{programming}&\textbf{economic}&{university}&{tennis}&{data}&{\textbf{home}}&\textbf{computer}&{golf}&{computers}&{stock}&{national}\\
    \hline
    \multicolumn{4}{|c|}{GPU-DMM}&\multicolumn{4}{|c|}{GPU-PDMM}&\multicolumn{4}{|c|}{BTM}\\
    \hline
    {Topic1}&{Topic2}&{Topic3}&{Topic4}&{Topic1}&{Topic2}&{Topic3}&{Topic4}&{Topic1}&{Topic2}&{Topic3}&{Topic4}\\
    {(health)}&{(politics)}&{(engine.)}&{(culture)}&{(health)}&{(politics)}&{(engine.)}&{(culture)}&{(health)}&{(politics)}&{(engine.)}&{(culture)}\\
    \hline
    {health}&{political}&{\textbf{theory}}&{movie}&{health}&{\textbf{wikipedia}}&{\textbf{com}}&{culture}&{health}&{political}&{car}&{movie}\\
    \textbf{information}&{culture}&{\textbf{theoretical}}&{music}&{gov}&\textbf{encyclopedia}&{\textbf{news}}&{art}&\textbf{information}&{\textbf{wikipedia}}&{engine}&{music}\\
    {cancer}&{democracy}&{\textbf{physics}}&\textbf{com}&{cancer}&{\textbf{wiki}}&{\textbf{information}}&{\textbf{american}}&{gov}&{culture}&{\textbf{intel}}&\textbf{com}\\
    {medical}&{\textbf{wikipedia}}&{\textbf{wikipedia}}&{film}&{medical}&{political}&{\textbf{home}}&{history}&{cancer}&{democracy}&{electrical}&{\textbf{amazon}}\\
    {gov}&{party}&{\textbf{edu}}&\textbf{news}&{disease}&{system}&{\textbf{online}}&{\textbf{car}}&{medical}&{\textbf{encyclopedia}}&\textbf{com}&{film}\\
    \textbf{news}&{system}&{\textbf{information}}&{movies}&{healthy}&{democracy}&{\textbf{world}}&{arts}&\textbf{news}&{party}&{digital}&\textbf{news}\\
    {healthy}&\textbf{information}&{\textbf{science}}&{\textbf{wikipedia}}&{nutrition}&{party}&{\textbf{web}}&{\textbf{imdb}}&{research}&{\textbf{wiki}}&{motor}&{movies}\\
    {disease}&{government}&{\textbf{research}}&{art}&{physical}&{government}&{\textbf{music}}&{museum}&{disease}&{system}&{wheels}&{books}\\
    {nutrition}&\textbf{news}&{\textbf{amazon}}&{fashion}&{hiv}&{gov}&{\textbf{index}}&{\textbf{income}}&{healthy}&{government}&{products}&{art}\\
    {hiv}&{gov}&{\textbf{com}}&{\textbf{amazon}}&{diet}&{war}&{\textbf{amazon}}&{literature}&{nutrition}&{war}&{automatic}&{video}\\
    \hline
    {Topic5}&{Topic6}&{Topic7}&{Topic8}&{Topic5}&{Topic6}&{Topic7}&{Topic8}&{Topic5}&{Topic6}&{Topic7}&{Topic8}\\
    {(sport)}&{(computer)}&{(business)}&{(education)}&{(sport)}&{(computer)}&{(business)}&{(education)}&{(sport)}&{(computer)}&{(business)}&{(education)}\\
    \hline
    {sports}&{computer}&{business}&{research}&{\textbf{movie}}&{computer}&{business}&{research}&{sports}&{computer}&{business}&{edu}\\
    \textbf{news}&{web}&{market}&{edu}&{sports}&{software}&{trade}&{edu}&\textbf{news}&{software}&\textbf{news}&{research}\\
    {football}&{software}&\textbf{news}&{science}&{games}&{programming}&{management}&{science}&{football}&{web}&{market}&{science}\\
    \textbf{com}&{programming}&{trade}&{school}&{football}&{systems}&{economic}&{theory}&\textbf{com}&{programming}&\textbf{information}&{theory}\\
    {games}&\textbf{com}&{stock}&{journal}&{\textbf{yahoo}}&{memory}&{law}&{school}&{games}&\textbf{com}&{trade}&\textbf{information}\\
    {soccer}&{wikipedia}&\textbf{information}&{university}&{game}&\textbf{engine}&{international}&{university}&{game}&{internet}&{stock}&{university}\\
    {game}&{memory}&\textbf{com}&{computer}&{\textbf{video}}&{intel}&{gov}&{journal}&{soccer}&{memory}&{services}&{journal}\\
    {\textbf{wikipedia}}&{intel}&{services}&\textbf{information}&{soccer}&{design}&{products}&{theoretical}&{match}&{data}&{\textbf{home}}&{school}\\
    {tennis}&{linux}&{finance}&{department}&{\textbf{film}}&{electrical}&{jobs}&{physics}&{world}&{wikipedia}&{gov}&{theoretical}\\
    {world}&{digital}&{\textbf{home}}&{graduate}&{\textbf{movies}}&{security}&{bank}&{department}&{tennis}&{linux}&{finance}&{physics}\\
    \hline
    \multicolumn{4}{|c|}{WNTM}&\multicolumn{4}{|c|}{SATM}&\multicolumn{4}{|c|}{PTM}\\
    \hline
    {Topic1}&{Topic2}&{Topic3}&{Topic4}&{Topic1}&{Topic2}&{Topic3}&{Topic4}&{Topic1}&{Topic2}&{Topic3}&{Topic4}\\
    {(health)}&{(politics)}&{(engine.)}&{(culture)}&{(health)}&{(politics)}&{(engine.)}&{(culture)}&{(health)}&{(politics)}&{(engine.)}&{(culture)}\\
    \hline
    {health}&{political}&{\textbf{music}}&\textbf{movie}&{health}&{\textbf{wikipedia}}&{\textbf{culture}}&\textbf{movie}&{health}&{\textbf{wikipedia}}&{\textbf{amazon}}&{music}\\
    \textbf{information}&{\textbf{wikipedia}}&\textbf{engine}&\textbf{com}&\textbf{information}&\textbf{research}&{\textbf{amazon}}&\textbf{news}&\textbf{information}&{\textbf{encyclopedia}}&{\textbf{theory}}&{movie}\\
    {cancer}&{culture}&\textbf{car}&{\textbf{amazon}}&\textbf{news}&\textbf{trade}&{\textbf{wikipedia}}&\textbf{com}&{cancer}&{\textbf{wiki}}&{\textbf{com}}&\textbf{com}\\
    {hiv}&{\textbf{encyclopedia}}&{\textbf{rock}}&\textbf{film}&{research}&\textbf{information}&{\textbf{democracy}}&\textbf{film}&{medical}&\textbf{political}&{\textbf{theoretical}}&\textbf{news}\\
    {healthy}&{system}&\textbf{com}&\textbf{art}&{gov}&\textbf{wiki}&{\textbf{books}}&\textbf{wikipedia}&{gov}&\textbf{democracy}&\textbf{physics}&{film}\\
    {nutrition}&{democracy}&\textbf{motor}&\textbf{books}&{medical}&\textbf{gov}&{\textbf{com}}&\textbf{movies}&\textbf{news}&\textbf{system}&\textbf{books}&{movies}\\
    {disease}&{party}&{\textbf{reviews}}&\textbf{fashion}&{\textbf{party}}&\textbf{international}&{\textbf{political}}&\textbf{reviews}&{disease}&\textbf{party}&\textbf{car}&{video}\\
    {medical}&{government}&{\textbf{pop}}&{\textbf{online}}&{\textbf{home}}&\textbf{journal}&{\textbf{history}}&\textbf{online}&{healthy}&\textbf{war}&\textbf{engine}&{reviews}\\
    {news}&{war}&{\textbf{band}}&\textbf{movies}&{disease}&\textbf{programming}&{\textbf{edu}}&\textbf{digital}&{nutrition}&\textbf{government}&\textbf{models}&{\textbf{intel}}\\
    {diet}&{world}&{\textbf{wikipedia}}&\textbf{video}&{healthy}&\textbf{business}&{\textbf{encyclopedia}}&\textbf{articles}&{hiv}&\textbf{house}&\textbf{electrical}&{imdb}\\
    \hline
    {Topic5}&{Topic6}&{Topic7}&{Topic8}&{Topic5}&{Topic6}&{Topic7}&{Topic8}&{Topic5}&{Topic6}&{Topic7}&{Topic8}\\
    {(sport)}&{(computer)}&{(business)}&{(education)}&{(sport)}&{(computer)}&{(business)}&{(education)}&{(sport)}&{(computer)}&{(business)}&{(education)}\\
    \hline
    {sports}&{computer}&{market}&{research}&{sports}&{system}&{business}&{edu}&\textbf{news}&{computer}&{business}&{research}\\
    {football}&{web}&{business}&{science}&{games}&\textbf{com}&\textbf{news}&{science}&{sports}&{edu}&\textbf{news}&{science}\\
    \textbf{news}&{programming}&{trade}&{edu}&\textbf{com}&{web}&\textbf{com}&{research}&\textbf{com}&{software}&\textbf{information}&{edu}\\
    {games}&{intel}&{stock}&{school}&\textbf{news}&{computer}&{market}&{school}&{football}&{web}&{market}&{journal}\\
    \textbf{com}&{memory}&\textbf{news}&{theory}&{football}&\textbf{information}&{yahoo}&\textbf{information}&{games}&{\textbf{school}}&{services}&{\textbf{art}}\\
    {soccer}&{internet}&\textbf{information}&\textbf{information}&{game}&{\textbf{car}}&{stock}&{university}&{game}&{research}&\textbf{com}&{resources}\\
    {game}&{systems}&{jobs}&{journal}&{world}&{wikipedia}&{internet}&{\textbf{program}}&{soccer}&{programming}&{trade}&{culture}\\
    {tennis}&\textbf{com}&{finance}&{university}&{soccer}&{memory}&{services}&{\textbf{fashion}}&{\textbf{culture}}&{\textbf{university}}&{stock}&\textbf{information}\\
    {match}&{data}&{\textbf{home}}&{theoretical}&{\textbf{online}}&{device}&{financial}&{\textbf{department}}&{world}&\textbf{information}&{\textbf{home}}&{directory}\\
    {league}&{wikipedia}&{tax}&{physics}&{\textbf{wikipedia}}&\textbf{engine}&\textbf{information}&{\textbf{home}}&{tennis}&{systems}&{gov}&{library}\\
    \hline
  \end{tabular}}
  
\end{table*}

\subsection{Influence of the number of iterations}

In this subsection, we try to investigate the influence of the number of iterations to the performance of all models using NMI metric. Since all models have converged when the number of iterations reaches 2000, we vary the number of iterations from 2 to 2024.

The results are shown in Figure \ref{NMIIter}. We can see that models based DMM can converge fast to the optimal solutions and almost get stable within 30 iterations. Models based global word co-occurrences get stable within 60 iterations. Models based self-aggregation has the slowest convergence speed and the lowest iterative performance. 

\begin{figure*}
\begin{minipage}{0.3\linewidth}
\centerline{\includegraphics[width=6cm]{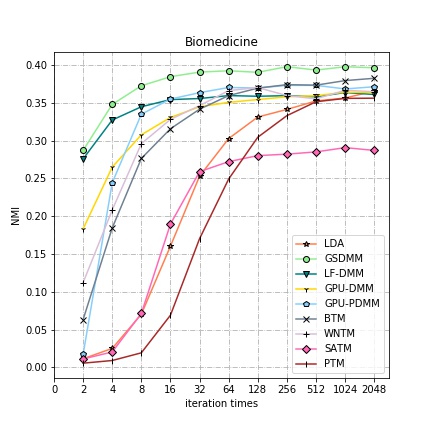}}
\centerline{(a). Biomedicine}
\end{minipage}
\hfill
\begin{minipage}{.3\linewidth}
\centerline{\includegraphics[width=6cm]{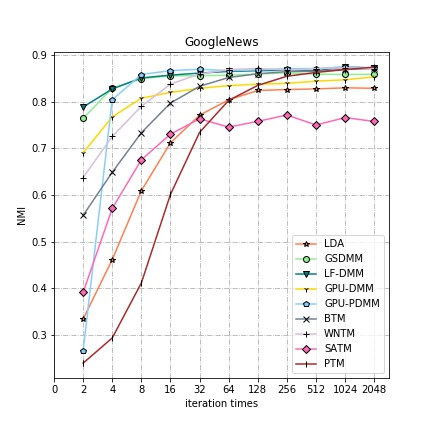}}
\centerline{(b). GoogleNews}
\end{minipage}
\hfill\begin{minipage}{0.3\linewidth}
\centerline{\includegraphics[width=6cm]{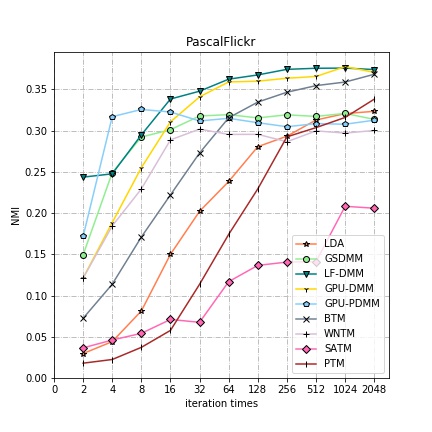}}
\centerline{(c). PascalFlickr}
\end{minipage}
\vfill
\begin{minipage}{.3\linewidth}
\centerline{\includegraphics[width=6cm]{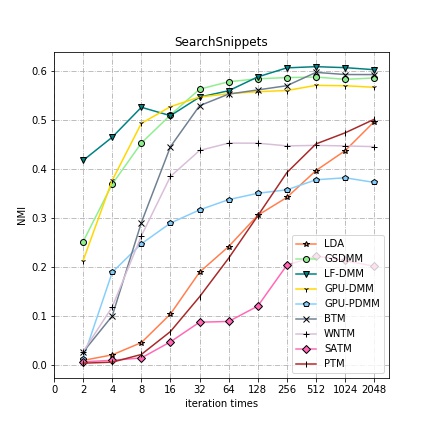}}
\centerline{(d). SearchSnippets}
\end{minipage}
\hfill
\begin{minipage}{0.3\linewidth}
\centerline{\includegraphics[width=6cm]{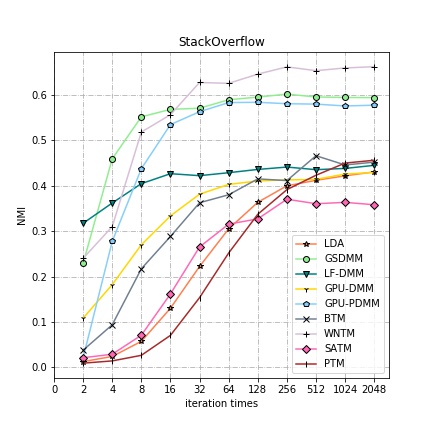}}
\centerline{(e). StackOverflow}
\end{minipage}
\hfill
\begin{minipage}{0.3\linewidth}
\centerline{\includegraphics[width=6cm]{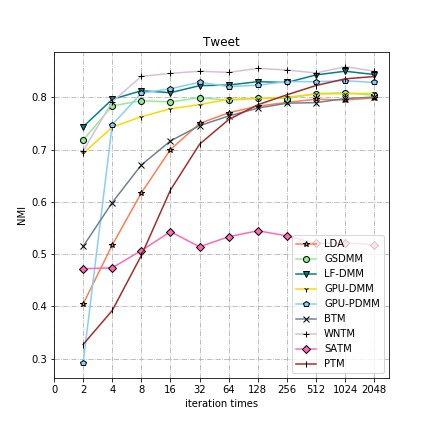}}
\centerline{(f). Tweet}
\end{minipage}
\caption{NMI values with different number of iterations on every corpora.} \label{NMIIter}

\end{figure*}

\subsection{Efficiency}

In this part, we compare the efficiency of various short text topic models. Here, we choose the largest dataset "Biomedicine" from all datasets to do the experiments. 

The average runtime of the initiation and per iteration for each model are reported in Table \ref{time}. Among all models evaluated, LDA and DMM are the most efficient methods as expected.  GPU-DMM is slightly slower than DMM and LDA, due to a similar Gibbs sampling process with GSDMM. LF-DMM and GPU-PDMM take much more time than GPU-DMM, because GPU-PDMM spends more time for the computational costs involved in sampling $\textbf{Z}_d$ and LF-DMM need much time for optimizing the topic vectors. We can see that GPU-PDMM is the slowest modeling compared with other models.

Global word co-occurrences based methods are much slower than GSDMM, LDA and GPU-DMM and faster than the rest models. This is expected since they extend the number of words by extracting word co-occurrences. For self-aggregation based methods, the time is affected by the number of pseudo-documents. PTM is much faster than SATM but much slower than global word co-occurrences based methods. In addition, the models by incorporating word embeddings (GPU-DMM, LF-DMM, and GPU-PDMM) have the slowest time for the initiation due to the computational cost for the similarity between words. 

\begin{table}[H]
  \centering
  \caption{The average runtime of initiation and per iteration of each model on Biomedicine (in milliseconds).} \label{time}
  \begin{tabular}{|l|l|l|}
    \hline
     {Model}&{Initiation time}&{Per iteration time}\\
     \hline
     {LDA}&77&41.50\\
     \hline
     {GSDMM}&46&48.15\\
     \hline
     {LF-DMM}&3329&2243.03\\
     \hline
     {GPU-DMM}&13610&53.83\\
     \hline
     {GPU-PDMM}&13037&9685.44\\
     \hline
     {BTM}&320&160.43\\
     \hline
     {WNTM}&192&220.12\\
     \hline
     {SATM}&41&2015.03\\
     \hline
     {PTM}&126&818.32\\
    \hline
  \end{tabular}
  
\end{table}

\section{Conclusion and Future Work}

The review of short text topic modeling (STTM) techniques covered three broad categories of methods: DMM based, global word co-occurrences based, and self-aggregation based. We studied the structure and properties preserved by various topic modeling algorithms and characterized the challenges faced by short text topic modeling techniques in general as well as each category of approaches. We presented various applications of STTM including content characterizing and recommendation, text classification, and event tracking. We provided an open-source Java library, named STTM, which is consisted of short text topic modeling approaches surveyed and evaluation tasks including classification, clustering, and topic coherence. Finally, we evaluated the surveyed approaches to these evaluation tasks using six publicly available real datasets and compared their strengths and weaknesses. 

Short text topic modeling is an emerging field in machine learning, and there are many promising research directions: (1) Visualization: as shown in the survey, we display topics by listing the most frequent words of each topic (see Figure \ref{top10}). This new ways of labeling the topics may be more reasonable by either choosing different words or displaying the chosen words differently \cite{chuang2012termite,sievert2014ldavis}. How to display a document using topic models is also a difficult problem? For each document, topic modeling provides useful information about the structure of the document. Binding with topic labels, this structure can help to identify the most interesting parts of the document. (2) Evaluation: Useful evaluation metrics for topic modeling algorithms have never been solved \cite{blei2012probabilistic}. Topic coherence cannot distinguish the differences between topics. In addition, existing metrics only evaluate one part of topic modeling algorithms. One open direction for topic modeling is to develop new evaluation metrics that match how the methods are used. (3) Model checking. From the experimental results on this paper, each method has different performance on different datasets. When dealing with a new corpus or a new task, we cannot decide which topic modeling algorithms should I use. How can I decide which of the many modeling assumptions are suitable for my goals? New computational answers to these questions would be a significant contribution to topic modeling.

\section*{Acknowledgement}

This research is partially supported by the National Natural Science Foundation of China under grant 61703362, and the Natural Science Foundation of Jiangsu Province of China under grant BK20170513.

\bibliographystyle{IEEEtran}
\bibliography{Survey}

\end{document}